\documentclass[11pt,a4paper]{article}
\usepackage{jcappub}
\usepackage{amsmath,amssymb,enumerate,epsfig,subfigure,caption2
}
\usepackage{overpic}
\usepackage{subfigure}
\usepackage[english]{babel}
\usepackage{amsmath}
\usepackage{amssymb}
\usepackage{epsfig}
\usepackage{graphics,psfrag,rotating}
\usepackage{graphicx}
\usepackage{dcolumn}
\usepackage{bm}%

\def\3nab{\tilde{\nabla}}

\def\be {\begin{equation}}
\def\ee {\end{equation}}
\def\bea {\begin{eqnarray}}
\def\eea {\end{eqnarray}}

\def\3nab{\tilde{\nabla}}

\def\hsp5{\hspace{5mm}}

\def\case#1/#2{\textstyle\frac{#1}{#2}}

\def\e{{\rm e}}

\def\bi {\bibitem}
\def\mc {\mathcal}
\def\case#1/#2{\textstyle\frac{#1}{#2} }

\newcommand{\n}{\hat{\nabla}}
\newcommand{\aD}{a_{\mc D}}
\newcommand{\HD}{H_{\mc D}}

\newcommand{\av}[1]{\left\langle#1\right\rangle_{\mc D}}
\def\diff{{\rm d}}

\title{Backreaction mechanism in multifluid and extended cosmologies}

\author{
Jose Beltr\'an Jim\'enez$^{a}\,\footnote{ jose.beltran [at] uclouvain.be}$, 
\'Alvaro de la Cruz-Dombriz$^{b}\, \footnote{ dombriz [at] fis.ucm.es}$ 
Peter K. S. Dunsby$^{c,d,e}\,\footnote{ peter.dunsby [at] uct.ac.za}$, 
Diego S\'aez-G\'omez$^{c,d,f}\footnote{ diego.saezgomez [at] uct.ac.za}$
}
\affiliation{
$^{a}$ Centre for Cosmology, Particle Physics and Phenomenology,
Institute of Mathematics and Physics, Louvain University,
2 Chemin du Cyclotron, 1348 Louvain-la-Neuve, Belgium.\\
$^{b}$Departamento de F\'{\i}sica
Te\'orica I, Ciudad Universitaria, Universidad Complutense de Madrid, E-28040 Madrid,
Spain.\\
$^{c}$  Astrophysics, Cosmology and Gravity Centre (ACGC), University of Cape Town, Rondebosch 7701, Cape Town, South Africa.\\
$^{d}$  Department of Mathematics and Applied Mathematics, University of Cape Town, Rondebosch 7701, Cape Town, South Africa.\\
$^{e}$  South African Astronomical Observatory,  Observatory 7925, Cape Town, South Africa.\\
$^{f}$ Fisika Teorikoaren eta Zientziaren Historia Saila, Zientzia eta Teknologia Fakultatea,\\
Euskal Herriko Unibertsitatea, 644 Posta Kutxatila, 48080 Bilbao, Spain \\
}

\date{\today}

\abstract{
One possible explanation for the present observed acceleration of the Universe is the breakdown of homogeneity and isotropy due to the formation of non-linear structures.  How inhomogeneities affect the averaged cosmological expansion rate and lead to late-time acceleration is generally considered to be due to some backreaction mechanism.  
In the recent literature most averaging calculations have focused their attention on General Relativity together with pressure-free matter.
In this communication we focus our attention on more general scenarios, including imperfect fluids as well as alternative theories of gravity, and apply an averaging procedure to them in order to determine possible backreaction effects. For illustrative purposes, we present our results for dark energy models, quintessence and Brans-Dicke theories. We also provide a discussion about the limitations of frame choices in the averaging procedure.}


\begin{document}
\maketitle


%
\section{Introduction}

The nature of the late time acceleration of the Universe \cite{Riess} remains an open and major problem in modern cosmology.
With the assumption of General Relativity (GR) as the correct gravitational theory, the standard Einstein field equations (EFE) when applied to sufficiently large scales, where the Universe is assumed to be homogeneous and isotropic, i.e., well-described by a Friedmann-Lema\^itre-Robertson-Walker (FLRW) model, give rise to decelerated periods of cosmological expansion whenever supplemented with either matter or radiation fluids. 
%
In fact, for the late-time evolution of the Universe, the GR predictions with standard (dust) matter break down by a factor of between one and two when confronted with observations 
\cite{Rasanen_Feb_11}. For instance, under the aforementioned assumptions the last scattering surface turns out to be larger and expansion rate longer than expected \cite{Refs_GR_wrong}.
Consequently, some approach to solving this problem is required in order to explain late-time cosmological acceleration. These approaches are usually classified in two different ways. The first -  and most popular one - considers that the Cosmological Principle assumption of homogeneous and isotropic spacetimes must be preserved at the expense of allowing the total stress-energy tensor appearing on the right-hand side (r.h.s.) of the EFE to be dominated at late times by a hypothetical negative pressure fluid usually dubbed dark energy (DE) \cite{Copeland}. An equivalent interpretation of this approach consists of modifying the left-hand side of EFE, thus modifying gravity itself, and interpreting the acceleration as a geometrical effect rather than as a consequence of the inclusion of non-physical fluids.
Both points of view are mathematically equivalent since geometrical modifications can be interpreted as curvature fluids and hence interpreted as DE contributions. Some examples of this include minimally-coupled models of scalar fields known as quintessence \cite{Quintessence} or more general K-essence models \cite{K-essence}, Lovelock theories \cite{Lovelock}, Gauss-Bonnet theories \cite{GB}, scalar-tensor theories like Brans-Dicke \cite{JFBD, Brans:2005ra, ST} or more general models \cite{generalST}, vector-tensor theories \cite{VT}, gravitational theories derived from extra dimensional models \cite{XD}; supergravity models \cite{sugra}, disformal theories \cite{disformal} or models  with either quantum-gravity-induced  violation or deformation of Lorentz symmetry and models of gravity breaking CPT \cite{LV}. In fact, the so-called $f(R)$ theories \cite{fR}, where the usual Einstein-Hilbert gravitational action is replaced by a more general $f(R)$ term, can be understood as a kind of scalar-tensor theory.

There are strong theoretical arguments to take into account scalar-tensor theories, including the fact that scalar partners of the graviton naturally arise in most attempts to quantise or unify gravity with other interactions and that the coupling between the scalar field and the matter density could provide a mechanism to alleviate the coincidence problem \cite{Chimento:2003iea}. Scalar-tensor theories are usually formulated in two different frames: the Jordan Frame (JF) and the Einstein Frame (EF). The former defines length and time as measured by standard laboratory apparatus, so that all observables (among others, time and redshift) have their standard interpretation in this frame. The metric is minimally coupled to matter in the JF and the scalar field is coupled to the Ricci curvature.  However, it is usually easier to perform calculations in the EF. This frame possesses the advantage that, in some simple cases, it diagonalises the kinetic terms for the spin-0 (the scalar field) and spin-2 (the graviton) degrees of freedom so that the presence of ghost, Laplacian and tachyonic instabilities can be directly identified. In this frame, the scalar field is coupled to matter \cite{Olive,Gilles}.


The second approach, which attempts to explain late-time acceleration considers that the cosmological homogeneity and isotropy assumptions, which are in fact statistical and coarse-grained, might be neglecting the possible influence that structure formation and subsequent growth of small-scale and non-linear structures may have on the cosmological expansion \cite{Clarkson-Ananda} or on light propagation \cite{Marozzi_Luminosity_redshift_relation}. The effects of astrophysical inhomogeneities on the averaged cosmological expansion is usually referred to as {\it backreaction} \cite{Ellis_Stoeger, Buchert:1995fz, Buchert:1999er} and has attracted a lot of attention in the last few years ({\it c.f.} \cite{Reviews} for reviews). 
The non-linear nature of GR or any other extended gravity theory ensures that the evolution for averaged fields
does not coincide with the evolution of inhomogeneous fields that are then averaged. Whether this difference is important or not is still a controversial matter ({\it c.f.} \cite{Chris-George-Obinna} for an extensive list of opinions on this subject).

The significance - if any - of backreaction can be understood as a consequence of non-Newtonian gravitational aspects beyond Newtonian theory and related to the differences between Newtonian gravity and the weak-field limit of GR \cite{Buchert:1999er,refs_Non_Newtonian}. This is an open question since the smallness of FLRW metric perturbations does not necessarily imply that averaged quantities remain close to the corresponding unperturbed values. In this regard, the authors in \cite{Rasanen_July2011} proved that provided the metric perturbations (and some of their derivatives) are small and the 4-velocity is close to its background value, then the redshift and averaged expansion remain close to the FLRW case, whereas the angular distance does not.
Simple models, with the inclusion of pressure-free matter have also demonstrated that accelerated expansion is possible \cite{Backreaction_simple_models} and how the distance-expansion rate relation turns out to be different from the FLRW case \cite{Rasanen_varia}. As an attempt to perform more realistic calculations, several proposals for observational constraints have started to become available \cite{Rockhee_et_al, Chris-George-Obinna}.
  
In order to determine the significance of backreaction, one possible approach, dubbed non-perturbative backreaction - contrarily to perturbative backreaction \cite{Paranjape:2009zu, Li-Thesis} -  consists of building a background model and its dynamics as a large-scale approximation of an inhomogeneous model. Results of such a process can be then compared with the hypothetical standard FLRW evolution. Several techniques are available 
\cite{Buchert:1995fz, Buchert:1999er, Rasanen_varia, Zalaletdinov1-2,Wiltshire1-2} in order to build up such a model. In this paper we will follow the so-called {\it  Buchert's approach} \cite{Buchert:1995fz, Buchert:1999er}.
The existing literature devoted to the backreaction mechanism and cosmological averaging usually focused on dust-dominated universes, geometrically described by GR,  although some generalisations to include a more general content were made in a gauge invariant way \cite{Marozzi_gauge_invariant} as well as some excursions into perfect fluid scenarios \cite{Buchert:2001sa}. The aim of this communication is therefore to extend the averaging procedure, and consequently the standard {\it Buchert's equations}, to general - imperfect - fluids when subject to averaging techniques and extract consequences for such a procedure. In particular, the impact that modifications of the integrability condition has on perturbative approaches will be addressed. As a natural consequence of our study, we will present the averaged equations for different extended theories of gravity and discuss the importance of frame choice in the interpretation of the averaged quantities.
%
%
This paper is organised as follows: in Section \ref{Section2} we present the kinematics and dynamics for general fluids that may encompass effective fluids arising from the new terms present in modified gravity theories. For such scenarios we derive in Section \ref{Section3} the generalised averaged equations. In this section we pay special attention to the generalised integrability condition.
Section \ref{Section4} is then devoted to applying the obtained formalism for several classes of modified gravity theories, where previous analysis done in Ref.~\cite{Vitagliano:2009zy} is fully extended. In this way we provide the relevant equations for a DE model with homogeneous equation of state, models of quintessence and finally for Brans-Dicke theories. 
In addition, Section \ref{EinsteinFrame} is devoted to applying the average procedure to the equations written in the Einstein frame, after a conformal transformation is applied, and a non-minimal coupling between the matter and the scalar field emerges. To conclude this section, we present a brief discussion about the limitations of the averaging procedure in the Einstein frame.
Finally, in Section \ref{conclusions} we present a brief discussion about the possible applications of our results and present the conclusions of this investigation.

Unless otherwise specified, we will use natural units ($\hbar=c=k_{B}=8\pi G=1$)
throughout this paper. Latin indices run from 1 to 3, whereas greek indices run from 0 to 3. The symbol $\nabla$ represents the 
usual covariant derivative, we use the
$(-,+,+,+)$ signature.  The Riemann tensor is defined by $
R^{\alpha}{}_{\beta\gamma\delta}=\Gamma^\alpha{}_{\beta\delta,\gamma}-\Gamma^\alpha{}_{\beta\gamma,\delta}+ \Gamma^\sigma{}_{\beta\delta}\Gamma^\alpha{}_{\gamma\sigma}-
\Gamma^\sigma{}_{\beta\gamma}\Gamma^\alpha{}_{\delta\sigma}$ where the $\Gamma^\alpha{}_{\beta\delta}$ are the Christoffel symbols defined by $
\Gamma^\alpha{}_{\beta\delta}=\frac{1}{2}g^{\alpha\sigma}
\left(g_{\beta\sigma,\delta}+g_{\sigma\delta,\beta}-g_{\beta\delta,\sigma}\right)$.
The Ricci tensor is obtained by contracting the {\em first} and the
{\em third} indices
$R_{\alpha\beta}=R^\mu_{\;\;\alpha\mu\beta}$.

\section{Formalism} \label{Section2}
In this section we present the general formalism for the decomposition of a general stress-energy tensor into its irreducible components. We then derive the fundamental local equations which are used in the subsequent sections to obtain the corresponding equations for the averaged quantities. 
\subsection{Kinematical quantities}
In order to decompose the stress-energy tensor and the evolution and constraint equations, let us use a time-like 4-vector $u^\mu$ and its associated orthogonal projection tensor $h_{\mu\nu}=g_{\mu\nu}+u_\mu u_\nu$, where $g_{\mu\nu}$ is the metric tensor on the full spacetime manifold. The set-up to keep in mind is a universe filled with a matter component plus an additional fluid\footnote{The additional fluid could represent contributions from dark energy, additional degrees of freedom or the effective effects associated with a modified theory of gravity.}. In this scenario, we will use the matter comoving frame and its associated 4-velocity to decompose all the kinematical quantities and equations. A crucial assumption that we will make is that matter is minimally coupled so that its flow is geodesic, i.e., $u^\mu\nabla_\mu u^\alpha=0$. This is not the case in some extended gravity theories described in certain frames, for example $f(R)$ or general Brans-Dicke theories in the Einstein frame \cite{Olive,Gilles}. This will be discussed in more detail below. 
Additionally, we will also assume that the matter flow is irrotational, which appears to be a well justified assumption on large enough scales \cite{vorticity}. Moreover,  having an irrotational flow will enable us to use its 4-velocity to globally foliate the spacetime on constant time hypersurfaces and set the time axis orthogonal to them and along the congruence \cite{Ellis-Cosmological Models}. In such cases the associated orthogonal projection tensor is the induced metric on the 3-hypersurfaces. In this  way, the metric for inhomogeneous and anisotropic universes can be expressed in terms of synchronous coordinates,
\begin{eqnarray}
{\rm d}s^2\,=\,-{\rm d}t^2+g_{ij}(t,\vec{x}){\rm d}x^i {\rm d}x^j
\label{metric_unperturbed}
\end{eqnarray}
where $i,j=1,2,3$ denote the spatial indices of the coordinate system in the 3-hypersurface. In this gauge, one can define comoving observers having $u^\mu=\delta^\mu_0$ and the projector tensor is nothing but the spatial part of the metric tensor $g_{ij}$. Moreover, we can adopt both 1+3 and 3+1 languages unambiguously and make contact with the geometry of foliations \cite{Gourgoulhon:2007ue}.

The kinematical quantities associated to the congruence described by $u^\mu$ are then the shear $\sigma_{\mu\nu}$ and the expansion $\theta$ so the covariant derivative of the congruence can be decomposed as
\begin{eqnarray}
\theta_{\mu \nu}\,\equiv\, \nabla_{\nu}u_{\mu}=\sigma_{\mu\nu}+\frac{1}{3}h_{\mu\nu}\theta.
\end{eqnarray}
Since our foliation is determined by $u^\mu$, we have that the extrinsic curvature $K_{\mu\nu}$ of the 3-hypersurfaces orthogonal to $u^\mu$ is precisely given by $K_{\mu\nu}=-\theta_{\mu\nu}$.

Some useful identities which will be used throughout this communication are\footnote{At this stage we should remind that our coordinates gauge choice makes that projected index onto the 3-hypersurface are precisely the spatial indices so that $\theta_{\mu\nu}$ being a completely projected tensor coincides with $\theta_{ij}$, and the same applies to the shear $\sigma_{\mu\nu}$}:
\begin{eqnarray}
\theta^{i}_{\,j}\,=\,\sigma^{i}_{\,j}+\frac{1}{3}\theta\delta^{i}_{\,j}\,\,\,;\,\,\, \theta^{i}_{\,j}\theta^{j}_{\,i}\,=\,2\sigma^2+\frac{1}{3}\theta^2\,\,\,;\,\,\,\sigma^{i}_{\;i}=0
\label{identities}
\end{eqnarray}
where the scalar shear is defined by $\sigma^2\equiv \frac12\sigma_{ij}\sigma^{ij}$. Angle brackets over indices will denote projected vectors onto the 3-hypersurfaces $v^{\langle \mu\rangle}\equiv h^\mu{}_{\nu}v^\mu$ and projected symmetric and trace free part of tensors $T_{\langle \mu\nu\rangle}\equiv (h^\alpha{}_{(\mu}h^\beta{}_{\nu)}-\frac13h_{\mu\nu}h^{\alpha\beta})T_{\alpha\beta}$. The 4-velocity also allows to define covariant derivatives along the congruence flow $\dot{T}\equiv\nabla_uT$ and covariant derivatives on the 3-hypersurfaces $\n_\mu \equiv h^{\alpha}{}_{\mu}\nabla_\alpha$, which is indeed a covariant derivative thanks to the absence of vorticity. Given our choice of frame and gauge, derivatives along the congruence will be simply time derivatives, i.e., $\dot{T}=\partial_t T$, which will play an important role in the averaging procedure.
\subsection{Stress-energy decomposition}
As commented above, our scenario will consist of a matter component plus an additional fluid so that the total stress-energy tensor will be given by $T_{\mu\nu}=T^m_{\mu\nu}+T^f_{\mu\nu}$. It is important to note that, since we are using the matter (dust) comoving frame we still have $T^m_{\mu\nu}=\rho_m u_\mu u_\nu$. However, the additional component will adopt the general form of a stress-energy tensor corresponding to an imperfect fluid so that it will be characterised by its energy density $\rho$, pressure $p$,  momentum flux $q_\mu$ and anisotropic stress $\pi_{\mu\nu}$, i.e., 
\be \label{EMT}
T_{\mu\nu}=\rho u_{\mu}u_{\nu}+ph_{\mu\nu}+2q_{(\mu}u_{\nu)}+\pi_{\mu\nu}\;.
\ee
The corresponding thermodynamical quantities are then given by 
\begin{eqnarray}
\rho&\equiv&T_{\mu\nu}u^{\mu}u^{\nu}\,\\
p&\equiv&\frac{1}{3}T_{\alpha\beta}h^{\alpha\beta}\,\\
q_\mu&\equiv&-T_{\alpha\beta}h^{\alpha}{}_{\mu}u^{\beta}
\,\\
\pi_{\mu\nu}&\equiv&T_{\alpha\beta}h^{\alpha}_{\langle \mu}h^{\beta}_{\nu\rangle}\;.
\end{eqnarray}
As usual, the momentum flux is orthogonal to the congruence $q_\mu u^\mu=0$ and the anisotropic stress has no components along the congruence $\pi_{\mu\nu} u^\nu=0$. If we take the covariant derivatives of these expressions one can easily show that $u^\mu\nabla_\nu q_\mu=-q^\mu\nabla_\nu u_\mu$ and $u_\nu\nabla_\mu\pi^{\mu\nu}=-\pi^{\mu\nu}\nabla_\nu u_\mu=-\pi^{\mu\nu}\sigma_{\mu\nu}$, which will be used below to simplify some expressions.
\subsection{Local propagation and constraint equations}
In order to proceed, the gravitational field equations will be written {\it \`a la Einstein} as
\begin{eqnarray}
G_{\mu\nu}\,&\equiv&\,R_{\mu\nu}-\frac{1}{2}g_{\mu\nu}R= T_{\mu\nu}= \sum_\alpha T^{(\alpha)}_{\mu\nu}\,,
\label{fieldtensorialequation_Einstein_manner}
\end{eqnarray}
where $T^{(\alpha)}_{\mu\nu}$ labels all possible contributions to the total stress-energy tensor, which reduces to two components in our case as explained above. 

The total stress-energy tensor in a general frame is determined by the density $\rho$, the pressure $p$, the momentum flux $q_\mu$ and the 
anisotropic stress $\pi_{\mu\nu}$. These contributions can be attributed to the presence of additional fields or as an effective description of a modified gravity theory.

Equipped with the geometrical quantities describing the kinematics of the fluids as well as the form of the more general fluid given by (\ref{EMT}), 
let us introduce the field equations in the so-called Arnowitt-Deser-Misner (ADM) decomposition:
\begin{eqnarray}
&&\frac{1}{2}\left(\mathcal{R}+\theta^{2}-\theta^{i}_{\,j}\theta^{j}_{\,i}\right)\,=\, \rho\,,
\label{const_eqn}\\
&&\theta_{,i}-\theta^{j}_{\;i;\,j}\,=\,8\pi G\, q_{i}\,,
\label{Spatial_constraint}
\\
&&\dot{\theta}^{i}{}_{j}=-\theta\theta^{i}{}_{j}-\mathcal{R}^{i}_{j}+\frac{1}{2} (\rho-3p)\delta^i{}_j + T^{i}_{\; j}\,,
\label{Evolution_eqn}
\end{eqnarray}
with $\mc R^i{}_j$ representing the spatial curvature of the 3-hypersurfaces orthogonal to the congruence and
the dot stands for derivatives with respect to proper (cosmic) time $t$ measured by comoving observers\footnote{Notice that given our choice of observer, the covariant derivative along the congruence coincides with the proper time derivative and, because of our gauge choice, it is given by $\partial_t$.}. The first two equations represent the constraint equations imposed by the lapse and the shift fields respectively, which are Lagrange multipliers corresponding to the invariance under diffeomorphisms. The last equation holds for the evolution equation provided by Einstein equations and, combined with the energy constraint yields the Raychauduri equation that determines the evolution of the expansion
\begin{eqnarray}
\dot{\theta}\,=\,-2\sigma^2-\frac{1}{3}\theta^2-\frac{1}{2}(\rho+3p)\,.
\label{Ray_eqn}
\end{eqnarray}
%
%
Finally we require an evolution equation for the shear scalar squared $\sigma^2$. In order to obtain it, it suffices to differentiate (\ref{identities}) and, after combining the result with (\ref{const_eqn}) and (\ref{Ray_eqn}), one gets
\begin{eqnarray}
\partial_t\sigma^{i}_{\;j}\,=\,-\theta\sigma^{i}_{\;j}-\mathcal{R}^{i}_{\perp\;j} + \pi^{i}_{\;j}
\end{eqnarray}
and consequently by contracting with the shear $\sigma^i_{\;j}$, the evolution equation for $\sigma^2$ yields
\begin{eqnarray}
\partial_t\sigma^2\,=\,-2\theta\sigma^2-\sigma^{i}_{\;j}\mathcal{R}^{i}_{\;j} + \sigma^{i}_{\;j}\pi^{j}_{\;i}
\end{eqnarray}
where the symbol  $\mathcal{R}^{i}_{\perp\;j}\equiv \mathcal{R}^{i}_{\;j}-\frac{1}{3}\mathcal{R}\delta^{i}_{\;j}$ has been introduced
in the last two equations.

In addition to the above decomposition of EFE, 
it will also be useful to use the conservation equation $\nabla_\mu T^{\mu\nu}=0$.  The projections of these equations 
along $u^\mu$ and onto the orthogonal hypersurfaces,  yield the following two identities
\begin{eqnarray}
-u_{\nu}\nabla_{\mu}T^{\mu\nu}\,&=&\,\dot{\rho}+\theta(\rho+p)+\nabla_{\mu}q^{\mu}+\pi^{\mu\nu}\sigma_{\mu\nu}=0\,,
\label{motion_eqn_0}
\\
h_{\alpha\nu}\nabla_{\mu}T^{\mu\nu}\,&=&\,\hat{\nabla}_{\alpha}p+\dot{q}_{\alpha}+\frac{4}{3}\theta q_{\alpha} + q^{\mu}\sigma_{\mu\alpha}+h_{\alpha\nu}\nabla_{\mu}\pi^{\mu\nu}\,=\,0\,,
\label{motion_eqn_i}
\end{eqnarray}
where we have used the fact that $u_{\nu}\nabla_{\mu}\pi^{\mu\nu}\,=\,-\pi^{\mu\nu}\sigma_{\mu\nu}$ as well as (\ref{identities}). 
At this stage let us define
\begin{eqnarray}
j_{}\,\equiv\,   
-\nabla_{\mu}q^{\mu}+\pi^{\mu\nu}\sigma_{\mu\nu}\,,
\label{jota_definition}
\end{eqnarray}
which will be useful later on. 
On the other hand, Eq. (\ref{motion_eqn_0}) can be obtained by combining the time derivative of (\ref{const_eqn}) with (\ref{Ray_eqn}) 
yielding 
\begin{eqnarray}
\dot{\rho}+\theta(\rho+p)\,=\,\left( \frac{1}{2}\dot{\mathcal{R}} +\frac{1}{3}\theta\, \mathcal{R} + \hat{\mathcal{R}}  \right)-\hat{\pi}\,,
\label{motion_eqn_0_v2}
\end{eqnarray}
with $\hat{\mathcal{R}}\equiv \sigma_{ij}\mathcal{R}^{ij}$ and analogously  $\hat{\pi}=\sigma_{ij}\pi^{ij}$. Terms in expressions (\ref{motion_eqn_0}) and (\ref{motion_eqn_0_v2}) 
can be matched using Codazzi-Gauss equations.


\section{Averaging procedure}
\label{Section3}
In this section we introduce the main definitions and relations of the averaging procedure. We then apply them to the local equations of the precedent section to obtain the corresponding averaged equations. In this way we obtain the averaged versions for Raychauduri and the continuity equations. Finally, we will obtain the so-called integrability condition relating the kinematical backreaction and the spatial average of the 3-curvature.
\subsection{Definitions and procedure}
In the literature, we can find several proposals on how to average an inhomogeneous spacetime. The main difficulty arises from the fact that for general tensors averaging is not a well-defined operation since it involves the evaluation of tensorial quantities at different spacetime points. In Buchert's approach \cite{Buchert:1999er}, this difficulty is circumvented by averaging only scalar quantities. We will use this approach in this work and will define the average of some scalar quantity $\mc O$ over some spatial region $\mc D$ as
\be
\av{\mc O}\equiv\frac{\int_{\mc D}{\mc O}\,\diff\Sigma}{\int_{\mc D}\,\diff\Sigma}
\label{average_definition}
\ee
with $\diff\Sigma$ the volume element on the spatial domain $\mc D$, which will be defined as a constant volume in comoving coordinates. Several choices can be found in the literature for the volume element of the domain corresponding to different measures. Here, we shall use the Riemannian measure. Since we are using the matter velocity geodesic congruence, no matter particles will cross the boundary of the domain and therefore, the total matter mass in the domain will remain constant. However, this does not imply a constant {\it total rest energy} within the domain, since additional components of the total stress-energy tensor might cross the domain boundaries. This is reflected by the fact that the total momentum flux $q^\mu$ is non-vanishing, so that $\int_{\mc D} q_\mu\diff \Sigma^\mu\neq0$, with $\diff\Sigma^\mu$ the volume element of the domain boundary. 

The Riemannian volume of the domain\footnote{In the Riemannian measure we use the determinant of the induced metric. However, one should be aware that the induced metric is the projector onto the 3-hypersurfaces and therefore, its determinant vanishes, since projector operators are not invertible. Thus, the induced metric must be understood as the projection tensor $h_{\mu\nu}$ evaluated on the surface. In practice, given our gauge choice, it is nothing but $g_{ij}$.} $V_{\mc D}=\int_{\mc D}\sqrt{h}\diff\Sigma$ allows one to define the effective scale factor $a_{\mc D}\equiv (V_{\mc D})^{1/3}$. Moreover, using the fact that $\theta$ can be related to the trace of the extrinsic curvature of the foliation via\footnote{Notice that this condition only holds for geodesic congruences since for an accelerated congruence the relation between $\theta_{\mu\nu}$ and the extrinsic curvature has a term proportional to $u_\mu a_\nu$.} $\theta=-h^{\mu\nu}K_{\mu\nu}$ and that the extrinsic curvature is nothing but $K_{\mu\nu}=-\frac12\partial_t h_{\mu\nu}$, we have that $\theta=-\partial_t{\sqrt{h}}/\sqrt{h}$. Thus, we can obtain the useful relation
\be
\av{\theta}=
\frac{1}{V_{\mc D}}\int_{\mc D}\theta\sqrt{h}\,\diff^3x=\frac{1}{V_{\mc D}}\int_{\mc D}\partial_t\sqrt{h}\,\diff^3x=\frac{\partial_tV_{\mc D}}{V_{\mc D}}
\ee
so that we finally obtain
\be
H_{\mc D}\equiv\frac{\partial_t a_{\mc D}}{a_{\mc D}}=\frac13\av{\theta},
\ee
which defines the effective Hubble expansion rate by means of the average of the congruence expansion.

From the averaging definition (\ref{average_definition}) it is also straightforward to prove the well-known $\it{commutation\,\, rule}$ for averaging and time derivative of a scalar quantity $\mathcal{O}$:
\begin{eqnarray}
\left[ \partial_{t},\,\av{}\right]\mathcal{O}\,\equiv\partial_t\av{\mc O}-\av{\partial_t\mc O}=\av{\theta\mc O}-\av{\theta}\av{\mc O},
\label{commutator}
\end{eqnarray}
which can also be expressed as
\be
\left[ \partial_{t},\,\av{}\right]\mathcal{O}=\av{\theta\delta\mc O}\,,
\ee
with $\delta \mc O\equiv \mc O-\av{\mc O}$. Now, if we use the fact that the average of $\delta\theta\equiv\theta-\av{\theta}$ vanishes (as for any perturbed scalar quantity), the commutator can be alternatively written in the form:
\be
\left[ \partial_{t},\,\av{}\right]\mathcal{O}=\av{\delta\theta\,\delta\mc O}\;.
\ee
This identity expresses the non-commutative character of the spatial averaging and time differentiating operations when applied to a given scalar quantity. Interestingly, the commutator vanishes when either the expansion or the scalar quantity do not differ from their mean value. In order words, time evolution and averaging commute when either $\theta$ or $\mc{O}$ are homogeneously distributed. If one of them is purely homogeneous, then the commutator vanishes. Of course, it is difficult to imagine a situation in which only one of them is homogeneous while the other one is inhomogeneous, since the inhomogeneities will be transferred to each other by means of the corresponding field equations\footnote{We will discuss this in more detail below within the context of quintessence or Brans-Dicke theories}. This is also related to the fact that the relevant quantity determining the level of non-commutavity is actually the correlation of the expansion perturbation and the perturbation of the scalar under consideration. Thus, effects from backreaction can only appear at second order in perturbations, which is the expected result, since backreaction can only become relevant due to the non-linearities of the equations. Therefore, this is the key fact explaining why inhomogeneities can acquire a relevant role in the averaged EFE and how, consequently, inhomogeneities can lead to observing global acceleration in a locally decelerating universe. This is of course a well-known fact and it is not the aim of the present work. 
Rather, we are interested in describing how this non-commutativity might affect the evolution of homogeneous cosmologies within the context of alternative gravity theories and/or DE models. This is indeed a crucial aspect of such scenarios since it is expected that the aforementioned effect will always be present and, even if it is proved to be small and irrelevant within the context of GR in a dust dominated universe, it could have important consequences in alternative scenarios.
\subsection{Averaged Einstein equations}
In the following, we apply the averaging procedure on the propagation and constraint equations obtained in the previous section in order to obtain the corresponding equations for the averaged quantities.
The spatial averaging procedure for Eqs. (\ref{const_eqn})-(\ref{Evolution_eqn}) together with the rule (\ref{commutator}) yields the effective Friedman equations
\begin{eqnarray}
H_{\mc D}^{2}\,&=&\,\frac{1}{3}\langle \rho \rangle_{\mc D}-\frac{1}{6}\langle \mathcal{R}\rangle_{\mc D}-\frac{1}{6}\langle Q \rangle_{\mc D}\;,
\label{eqn_averaged_1}
\\
\frac{\ddot{a}_{\mc D}}{a_{\mc D}}\,&=&\,-\frac{1}{6}\langle \rho+3p\rangle_{\mc D}+\frac{1}{3}\langle Q \rangle_{\mc D}\;,
\label{eqn_averaged_2}
\\
\partial_{t}\langle \sigma^2 \rangle_{\mc D}\,&=&\,-2\langle \theta \rangle_{\mc D} \langle \sigma^2 \rangle_{\mc D} - 
 \langle \theta\,\delta \sigma^2\rangle_{\mc D}
+\langle\sigma^{i}_{\;j} \mathcal{C}^{j}_{\;i} \rangle_{\mc D},
\label{eqn_averaged_3}
\end{eqnarray}
where $H_{\mc D}\equiv\dot{a}_{\mc D}/a_{\mc D} = \langle \theta \rangle_{\mc D}/3$.
Note that these equations can be interpreted as a generalisation of the standard Buchert equations since a more general fluid as given in (\ref{EMT}) has been introduced. 
Thus, let us refer to the set of equations
(\ref{eqn_averaged_1}) - (\ref{eqn_averaged_3})
as generalized Buchert equations.
In the previous equations, the following definitions have been introduced
\begin{eqnarray}
 \langle Q \rangle_{\mc D}\,&\equiv&\, \frac{2}{3}\left(  \langle \theta^2 \rangle_{\mc D}^{}- \langle \theta \rangle_{\mc D}^{2}   \right)-2 \langle \sigma^2 \rangle_{\mc D}\,,
\label{def_eqn_averaged_1}
\\
 \delta\sigma^2
&\equiv&\, \sigma^2-\av{\sigma^2}\,,
\label{def_eqn_averaged_2}
\\
\mathcal{C}^{j}_{\;\;i}\,&\equiv&\,  \pi^{j}_{\;i}-\mathcal{R}^{j}_{\perp\;\;i}\,,
\label{def_eqn_averaged_3}
\end{eqnarray}
where $\langle Q\rangle_{\mc D}$ is usually referred to as the {\rm kinematical backreaction term}. As we can see from Eq. (\ref{eqn_averaged_1}), this term contributes to the averaged expansion as an effective additional fluid and this is why it has been suggested as a possible explanation for Dark Energy. Moreover, as we can conclude from its definition, the kinematical backreaction becomes more important as the expansion gets more inhomogeneous, which could eventually solve the coincidence problem because $\av{Q}$ starts being relevant when structures start forming.
\subsection{Averaged continuity equation and integrability condition}
We now obtain the averaged version of the continuity equation that will allow us to obtained the generalised integrability condition relating the kinematical backreaction and the average of the spatial scalar curvature. The averaging procedure applied to the continuity equation (\ref{motion_eqn_0}) yields the following equation
\begin{eqnarray}
\partial_{t} \av{ \rho}+\av{\theta} \av{ \rho}+ \av{\theta p}\,=\, \av{j}
\end{eqnarray}
which obviously reduces to the standard averaged conservation of mass when only a dust fluid with $p=j=0$ is considered. If we now use the fact that 
\begin{eqnarray}
 \langle \theta p \rangle_{\mc D} \,=\, \langle \theta \rangle_{\mc D}  \langle p \rangle_{\mc D}     + \langle \theta \left( p-  \langle p \rangle_{\mc D} \right) \rangle_{\mc D}\,=\,    
 \langle \theta \rangle_{\mc D}  \langle p \rangle_{\mc D}     + \langle \theta \delta p \rangle_{\mc D}
\end{eqnarray}
together with the vanishing of the averaged pressure perturbation $\av{\delta p}=0$,
the continuity equation can be alternatively written in the more familiar form
\begin{eqnarray}
\partial_{t} \av{ \rho}+\av{\theta} \av{ \rho+p}\,=\, -\av{\delta\theta\,\delta p}+\av{j}
\label{continuity_averaged}
\end{eqnarray}
where one can clearly see the backreaction effects as a source coming from the non-commutativity of time evolution and averaging as well as the term $\av{j}$.
Now that the averaged continuity equation has been obtained, we can combine equations \eqref{commutator}, 
\eqref{eqn_averaged_1}, \eqref{eqn_averaged_2} and \eqref{continuity_averaged}
to obtain a generalized integrability condition relating $\langle Q\rangle_{\mc D}$ and $\langle \mc{R}\rangle_{\mc D}$ as follows
\begin{eqnarray}
\frac{1}{2 a_{\mc D}^6}\Big[\partial_{t} \left(a_{\mc D}^6 \av{Q } \right)+a_{\mc D}^4\partial_{t}\left(a_D^2 \av{\mathcal{R}} \right)
\Big]\,=\,
- \av{\delta\theta \, \delta p } + \av{j}.
\label{generalized_continuity_averaged}
 \end{eqnarray}
This equation
has no analogue in Newtonian dynamics or even in the case of GR.  The two terms on the 
r.h.s. of (\ref{generalized_continuity_averaged}) are absent in dust universes geometrically 
described by GR. In this special case, (\ref{generalized_continuity_averaged}) 
becomes the usual integrability condition in standard averaging (see for instance 
equation (13b) in \cite{Buchert:1999er}). However, either in the case of an arbitrary 
fluid, or modified gravity, the terms on the r.h.s. are non-zero, leading to very different behaviour with respect to GR when averaging is applied. 

For instance, in the frame of perturbative backreaction \cite{Paranjape:2009zu, Li-Thesis}, 
the non-zero contribution in the r.h.s. of (\ref{generalized_continuity_averaged}) would 
render calculations of the second-order contributions for the averaged spatial curvature $\av{\mathcal{R}}$ more tortuous. In fact,  unlike the standard GR and dust universe case, 
it would not suffice to know the kinematical backreaction 
$\av{Q}$\footnote{$\av{Q }$ can be proved to include only second-order contributions depending its expression only upon squares of first-order terms.} 
in order to determine directly the second-order contributions for $\av{\mathcal{R}}$ precisely due to the terms on the r.h.s. of 
 (\ref{generalized_continuity_averaged}). 
Since the integrability condition  is an exact relation valid to any order, calculations may be ultimately performed but at the expense of altering the standard procedure.

Another interesting novelty in this scenario is that in cases where the scalar curvature averages as that of a FLRW universe with $\av{\mc R}\propto a_{\mc D}^{-2}$ (vanishing r.h.s. of the integrability condition), is that the kinematical backreaction must evolve as $\av{\mc Q}\propto a_{\mc D}^{-6}$, so it becomes quickly diluted as the universe expands. However, in our alternative scenario with the presence of an additional (effective) fluid, the integrability condition (\ref{generalized_continuity_averaged}) has a non-vanishing r.h.s. so that, even if the spatial curvature averages as that of a FLRW metric, the kinematical backreaction will still be sourced by the pressure perturbation, the momentum flux and the anisotropic stress so that it does not need to become quickly diluted. Even if the additional fluid behaves like a perfect fluid but has pressure perturbations, it will source the kinematical backreaction, as we shall see in more detail in next section.

Finally, it is worth mentioning the well-known fact that the averaged equations do not form a closed system so that some assumptions need to be made in order to compute the  phenomenological consequences. 
\section{Averaging in extended cosmologies}
\label{Section4}
In this section we apply the results obtained in the two previous sections to three classes of extended gravity theories: a dark energy model effectively described as a perfect fluid, a quintessence model and Brans-Dicke theories. We should remind one that in what follows we shall assume the standard matter fluid to be perfect, i.e., $q_{m\;\mu}=0$ and $\pi_{m\;\mu\nu}=0$. 
\subsection{Perfect fluid dark energy}
\label{Subsection_DE}
%
%
Let us start by studying a universe with matter plus a dark energy component effectively characterized by a perfect fluid whose energy-momentum tensor is given by
\be
T^{\rm DE}_{\mu\nu}=(\rho_{\rm DE}+p_{\rm DE})v_\mu v_\nu+p_{\rm DE}g_{\mu\nu}.
\ee
It is important to bear in mind that, in general, the comoving frame of dark energy will be different from that of the matter component and, thus, $v_\mu$ will not necessarily coincide with the matter flow $u_\mu$ that we are using for our foliation. In models for dark energy in the form of a perfect fluid is usually assumed that both matter and dark energy have the same background rest frame, i.e., that $v^{(0)}_\mu(t)=u^{(0)}_\mu(t)$ at zeroth order and differences only appear as peculiar velocities. However, as argued in \cite{movingDE}, if dark energy was always decoupled from matter and the rest of components of the universe, there is no reason a priori to expect that both components will share a common rest frame even on the largest scales and a net coherent flow between them might exist. In any case, this coincidence of rest frames cannot be maintained at the level of perturbations and since we are precisely dealing with the inhomogeneous case, we need to be careful about this fact and take it into account by letting $v_\mu\neq u_\mu$. If the difference between both frames is characterised by $w_\mu=v_\mu-u_\mu$, the dark energy stress-energy tensor reads
\be
T^{\rm DE}_{\mu\nu}=
(\rho_{\rm DE}+p_{\rm DE})u_\mu u_\nu+p_{\rm DE}g_{\mu\nu}
+2(\rho_{\rm DE}+p_{\rm DE})u_{(\mu}w_{\nu)}+(\rho_{\rm DE}+p_{\rm DE})w_\mu w_\nu.
\ee
As expected, even though dark energy is described by a perfect fluid, its stress-energy tensor acquires momentum flux and anisotropic stress contributions when expressed in the matter rest frame. It is also not surprising that for a cosmological constant-like fluid with $p_{\rm DE}=-\rho_{\rm DE}$, the momentum flux and anisotropic stress vanish also in the matter rest frame.

Then, EFE read
\begin{eqnarray}
G_{\mu\nu}\,=\,8\pi G\Big[\left(\rho_{m}+\rho_{{\rm DE}}+p_{{\rm DE}}\right)u_\mu u_\nu+p_{{\rm DE}}\,g_{\mu\nu}+2q_{(\mu}^{\rm DE}u_{\nu)} +\pi^{\rm DE}_{\mu\nu}\Big],
\label{DE_model_eqns}
\end{eqnarray}
with $\nabla_{\mu}T^{\mu\nu}_{m}=0=\nabla_{\mu}T^{ \mu\nu}_{\rm DE}$, i.e., the DE component does not interact with standard matter. Note that our congruence defining the time-like direction has been chosen to coincide with the matter flow, i.e., $u^\mu$ corresponds to the comoving frame with matter. Since matter is a pressureless fluid, one can have synchronous coordinates as the ones we are using because no pressure gradients will be generated for matter.

The corresponding averaged continuity equations for the two fluids become
\begin{eqnarray}
\partial_{t} \av{\rho_{\rm m}}  + \av{\theta} \av {\rho_{\rm m}} \,&=&0\\
\partial_{t} \av{ \rho_{\rm DE}}+\av{\theta} \av{ \rho_{\rm DE}+p_{\rm DE}}\,&=&\, -\av{\delta\theta\,\delta p_{\rm DE}}+\av{j_{\rm DE}}.
\end{eqnarray}
where the relation
$j_{\rm DE}\,\equiv\,  -\nabla^{\mu}q^{\rm DE}_{\mu}+\pi^{\rm DE}_{\mu\nu}\sigma^{\mu\nu}$
according to $(\ref{jota_definition})$ has been used.
At this stage let us mention that the continuity equation $(\ref{motion_eqn_i})$ renders a non trivial relation between the momentum flux and the spatial gradients of dark energy pressure. The orthogonal projection of the dark energy stress-energy conservation equations onto the 3-hypersurfaces is given by
\begin{eqnarray}
\hat{\nabla}_{\alpha}p_{\rm DE}+\dot{q}^{\rm DE}_{\alpha}+\frac{4}{3}\theta q^{\rm DE}_{\alpha} 
+ q^{\rm DE}_{\mu}\sigma^\mu_\alpha+h_\alpha^\nu\nabla^{\mu}\pi^{\rm DE}_{\mu\nu}
\,=\,0
\end{eqnarray}
showing the relation between non-vanishing momentum flux and anisotropic shear for dark energy and the existence of dark energy pressure gradients. In the most general case, there would be an acceleration term in this equation signalling that, even for a general perfect fluid in its rest frame, pressure gradients will deviate the fluid flow from being geodesic. However, it is worth stating once again that we are using the matter rest frame, so the acceleration of the frame is zero.  It follows that for the dark energy component, we can see that pressure gradients are supported by momentum flux and anisotropic stresses. 
The integrability condition simplifies to
\begin{eqnarray}
\frac{1}{2 a_{\mc D}^6}\Big[\partial_{t} \left(a_D^6 \av{Q}  \right)+a_D^4\partial_{t}\left(a_D^2 \av{\mc R}\right)
\Big]\,=\,
- \av{\delta\theta\,\delta p_{\rm DE}}+\av{j_{\rm DE}}.
\label{Integrability_Condition_DE}
\end{eqnarray}
Notice that the r.h.s of the commutation relation only depends on dark energy quantities because the matter fluid is pressureless and we are using its rest frame. In both the dark energy continuity equation and commutation relations, the backreaction corrections are given in terms of the pressure perturbation of the DE component. In general, the pressure perturbation will encompass both an adiabatic contribution determined in terms of its adiabatic sound speed $c_s^2$ and an entropic contribution. If we assume adiabaticity, the sound speed fully determines the clustering properties of the fluid \footnote{See however \cite{Ballesteros:2010ks} for a careful discussion about this point and the effects of DE with non-adiabatic sound speed. It is worth noting that the sound speed defined as $\delta p/\delta \rho$ is a gauge-dependent quantity, but the well-defined and gauge independent sound speed is  $\delta p/\delta \rho\vert_{\rm rest frame}$, i.e., the sound speed evaluated in the rest frame of the fluid and this is indeed the frame that we are using. }. In most of the DE models the sound speed is close to 1, which makes its Jeans' scale larger than the Hubble scale so that DE clustering within the horizon does not occur. The underlying reason for this is that pressure will prevent gravitational collapse from being efficient and, therefore, the formation of structures. For these models, the corrections coming from backreaction are expected to be very small since they represent a second order correction with a very small perturbation on all sub-Hubble scales. However, certain extensions of the simplest DE models allow for a Jeans scale significantly smaller than the Hubble scale, like the K-essence models, so that DE can undergo a clustering process that might lead to non-trivial backreaction effects on the averaged evolution. In the previous discussion we have assumed adiabaticity, so that the sound speed fully characterises the DE perturbations. Of course, in more general frameworks with non-adiabatic perturbations or even imperfect fluids \cite{Sawicki:2012re}, the conclusions will change and a more detailed study is required in order to determine the precise consequences extracted from the averaging procedure and how backreaction might affect the cosmological evolution in such cases. 
A common characterisation of the DE fluid is by means of
a barotropic equation of state of the form  $p_{{\rm DE}}=\omega\rho_{{\rm DE}}$. For an arbitrary equation of state parameter we have $ \av{ p_{\rm DE} }  \neq \omega \av{ \rho_{{\rm DE}} }$ unless $\omega$ is homogeneous. More specifically, one finds $
 \av{p_{\rm DE}}=\av{w}\av{\rho_{\rm DE}}+\av{w\delta\rho_{\rm DE}}$ or, equivalently $\av{p_{\rm DE}}=\av{w}\av{\rho_{\rm DE}}+\av{\delta w\,\rho_{\rm DE}}$. Again, using the vanishing of the average perturbed quantities $\av{\delta w}=\av{\delta\rho_{\rm DE}}=0$ we can alternatively write
 \be
 \av{p_{\rm DE}}=\av{w}\av{\rho_{\rm DE}}+\av{\delta w\delta\rho_{\rm DE}}
 \ee
where we can clearly see how the averaged equation of state acquires a second order correction when the equation of state parameter and the energy density are both inhomogeneous. For the sake of simplicity and as commonly considered in the literature, we shall now study the case with homogeneous $w$, which can even be just a constant (for sufficiently low redshifts and close to $-1$) as it happens for the standard matter and radiation components. Under this assumption we have $ \av{ p_{\rm DE} }=\omega \av{\rho_{\rm DE}}$ and the continuity equation (\ref{continuity_averaged}) now reads
 \begin{eqnarray}
\partial_{t} \av{ \rho_{\rm DE}}+3(1+w)H_{\mc D} \av{ \rho_{\rm DE}}\,&=&\, -\av{\delta\theta\,\delta p_{\rm DE}-j_{\rm DE}}\,,
\end{eqnarray}
where we have used that $\av{\theta}=3 H_{\mc D}$. This equation resembles the usual continuity equation for a homogeneous perfect fluid with an external source, that might be interpreted as an interaction with the perturbations. If we assume constant equation of state parameter, the continuity equation can be formally solved by\footnote{It is also straightforward to obtain the analogous solution for the case with $w=w(t)$, but we prefer to neglect its possible time-dependence for simplicity.}
\be
\av{\rho_{\rm DE}}=\rho_{\rm DE}^0\, a_{\mc D}^{-3(1+w)}\left[1-\int_{a^{0}_{\mc D}}^{a_{\mc D}} \tilde{a}_{\mc D}^{3(1+w)}\av{\delta\theta\,\delta p_{\rm DE}-j_{\rm DE}}\frac{\diff \tilde{a}_{\mc D}}{\tilde{a}_{\mc D}\tilde{H}_{\mc D}}\right].
\ee
From this expression we can conclude that backreaction effects can {\it redress} the equation of state parameter for DE. There is even the potential effect of having an {\it effective} phantom dark energy equation of state, while the bare $w$ is perfectly above the phantom divide line $w=-1$.
  In this regard, the possibility of dressing the cosmological parameters by means of backreaction effects was already suggested in \cite{Buchert:2002ij}. Also in \cite{DE_Marozzi}, the effects of stochastic perturbations on the dark energy parameters were analysed and shown that could lead to statistical variations of a few percent in the determination of the dark energy density parameter. 

If we assume a power-law evolution for the backreaction {\it source} term of the continuity equation\footnote{As commented above, the system of averaged equations do not form a closed system. Here we get this difficulty around by assuming a specific form of some averaged quantities. Although the full validity or our ansatz should be justified on more theoretical or observational grounds, we feel that a power law evolution, being a common behaviour in cosmological scenarios, is a quite reasonable assumption and sufficient for our illustrative purposes.}, i.e., $\av{\delta  \theta\, \delta p_{\rm DE}-j_{\rm DE}}\propto a^{m}_{\mc D}$, we can write
\be
\av{\rho_{\rm DE}}=\rho_{\rm DE}^0 a_{\mc D}^{-3(1+w)}\left[1+A\aD^{3(1+w)+m-p}\right]
\label{eq_rho_DE_averaged}
\ee
where $A$ is some constant amplitude and we have also assumed that $\HD\propto \aD^p$. Thus, the backreaction correction in (\ref{eq_rho_DE_averaged}) will be increasing during the expansion of the universe, provided that $3(1+w)+m-p>0$. For a slowly rolling scalar field with $w\simeq-1$, the condition is approximately given by $m>p$ or, equivalently, $\av{\delta\theta\, \delta p_{\rm DE}-j_{\rm DE}}$ grows faster than the averaged expansion $\HD$.

Let us finish our discussion of this scenario coming back to the  
 integrability condition (\ref{Integrability_Condition_DE}) 
%
%
%
As already discussed, this integrability condition is modified with respect to the standard GR result in the presence of a matter component so that, even if the spatial curvature averages as that of a FLRW metric, the kinematical backreaction $\av{\mc Q}$ does not need to decay as $\aD^{-6}$. In fact, under the same assumption as before, i.e., $\av{\delta\theta\, \delta p_{\rm DE}-j_{\rm DE}}\propto\aD^{m}$, and $\HD\propto\aD^p$, the integrability condition yields
\be
\av{Q}=C_1\aD^{m-p}+C_2\aD^{-6}
\label{averaged_Q}
\ee
with $C_{1,2}$ some constants. We clearly see how the kinematical backreaction differs from the standard result given by the mode $C_2$. As in the averaged DE density, the correction in (\ref{averaged_Q}) is determined by the ratio $(\av{\delta\theta\, \delta p_{\rm DE}-j_{\rm DE}})/\HD$ so that, whenever this ratio grows, the kinematical backreaction becomes more important as the universe expands.
%
%
\subsection{Quintessence}
\label{Subsection_Quintessence}
we will now consider a field theory model of DE based on a single scalar field minimally coupled to gravity and with a given potential. The total action for such theories supplemented with the usual Einstein-Hilbert term can be written as \cite{Quintessence}
\begin{eqnarray}
S\,=\,\int{\rm d}^4x \sqrt{-g}\left(\frac{1}{2}R - \frac{1}{2}\partial_\mu \phi\partial^\mu\phi-V(\phi) \right).
\label{actionQ}
\end{eqnarray}
The EFE for these theories are given by
\begin{eqnarray}
G_{\mu\nu}\,=\,\Big(T^{(m)}_{\mu\nu}+T^{Q}_{\mu\nu}\Big),
\label{Qessence_EFE}
\end{eqnarray}
where the quintessence field energy-momentum tensor and the field equation of motion become
\begin{eqnarray}
T_{\mu\nu}^{Q}\,=\, \partial_{\mu} \phi \,\partial_{\nu} \phi - g_{\mu\nu}\left[\frac{1}{2}\left(\partial\phi\right)^2+V(\phi)\right]\,,
\label{EM-Quintessence}
\end{eqnarray}
and 
\begin{eqnarray}
\Box \phi\,=\, \frac{\diff V(\phi)}{\diff \phi}\,,
\end{eqnarray}
respectively. Thus, whenever the potential is flat enough so that the field is slowly rolling down, the scalar field can drive a period of accelerated expansion. 

The corresponding thermodynamical quantities for quintessence theories, including standard matter, become
\begin{eqnarray}
\label{mu_Q}
\rho\,&\equiv\,&	T_{\mu\nu}u^{\mu}u^{\nu}\,=\,\rho_{m}+\frac{1}{2}\dot{\phi}^2+\frac{1}{2}h^{\alpha\beta}\hat{\nabla}_{\alpha}{\phi}\hat{\nabla}_{\beta}+V(\phi)
\\
%
\label{p_Q}
p\,&\equiv&\,\frac{1}{3}T_{\mu\nu}h^{\mu\nu}\,=\,p_{m}+\frac{1}{2}\dot{\phi}^2-\frac{1}{6}h^{\alpha\beta}\hat{\nabla}_{\alpha}{\phi}\hat{\nabla}_{\beta}-V(\phi)
\\
%
\label{q_Q}
q_{\mu}\,&\equiv&\,-T_{\alpha\beta}h^{\alpha}_{\mu}u^{\beta}\,=\,-\dot{\phi}\hat{\nabla}_{\mu}\phi
\\
\label{pi_Q}
\pi_{\mu\nu}\,&\equiv&\, T_{\alpha\beta}h^{\alpha}_{\;<\mu}h^{\beta}_{\;\nu>}    \,=\,\hat{T}_{\mu\nu}-p h_{\mu\nu}\;\;\;,\;\;\;  \hat{T}_{\mu\nu}\,\equiv\,h^{\alpha}_{(\mu}h^{\beta}_{\nu)}T_{\alpha\beta}
\end{eqnarray}
where again the matter fluid has been assumed to be perfect, i.e., $q_{m\;\mu}=0$ and $\pi_{m\;\mu\nu}=0$. 
In the following we shall study the backreaction effects on the evolution of the scalar field on a given background metric. To that end, we shall first decompose the scalar field equation into covariant derivatives along the congruence $u^\mu$ and those corresponding to the orthogonal 3-hypersurfaces as follows:
\be
\ddot{\phi}+\theta\dot{\phi}-h^{\alpha\beta}\n_\alpha\n_\beta\phi+V_{,\phi}=0
\ee
If we take the average of this equation and make repeated use of the commutation relation, we obtain 
\be
\partial_{tt}\av{\phi}+\av{\theta}\partial_t\av{\phi}+\av{V_{,\phi}}=\partial_t\av{\delta\theta\,\delta\phi}+\av{\theta}\av{\delta\theta\,\delta\phi}+\av{h^{\alpha\beta}\n_\alpha\n_\beta\phi}.
\label{Quintessence_averaged}
\ee
In quintessence models, DE is ascribed to the evolution of the homogeneous scalar field. This actually means neglecting all the terms on the r.h.s. of the above equation for the field evolution. However, we can see that inhomogeneous perturbations will source the homogeneous evolution through the r.h.s of this equation. 
If we look at the averaged equation (\ref{Quintessence_averaged}), we see that the mean value of $\phi$ will evolve in the same manner as the homogeneous mode if and only if the terms on the r.h.s of this equation 
are negligible. This is indeed the condition for the consistency of considering a pure homogeneous field, since only under such circumstances one can guarantee that $\phi(t)$ evolves in the same way as $\av{\phi}$. An alternative way of rewriting (\ref{Quintessence_averaged}) yields
\be
\partial_{t}\Big(\partial_t\av{\phi}-\av{\delta\theta\,\delta\phi}\Big)+\av{\theta}\Big(\partial_t\av{\phi}-\av{\delta\theta\,\delta\phi}\Big)+\av{V_{,\phi}}=\av{h^{\alpha\beta}\n_\alpha\n_\beta\phi}.
\label{Quintessence_averaged_bis}
\ee
In the usual case of quintessence models without taking care of the averaging, the field remains approximately frozen as long as its mass (determined by the potential) is smaller than the expansion. However, provided one ignores the r.h.s of 
(\ref{Quintessence_averaged_bis}), i.e., one assumes a homogeneous field, and the potential is much smaller than the expansion (slow-roll condition), what one actually finds is
\be
\partial_t\left[\aD^3\Big(\partial_t\av{\phi}-\av{\delta\theta\,\delta\phi}\Big)\right]=0\,
\ee
where we have used that $\av{\theta}=3\partial_t\aD/\aD$. Thus, the averaged field evolves as
\be
\av{\phi}\simeq\phi_0+\int\frac{\av{\delta\theta\,\delta\phi}}{\aD\HD}\diff\aD\,,
\ee
where $\phi_0$ is the usual constant mode and we have neglected the decaying mode\footnote{The mode we are neglecting evolves as $\int\aD^{-3}\diff t$ which can be a growing mode in some scenarios, but decays for the usual radiation and matter dominated epochs.}. As was also found in the previous section, the evolution of the averaged field possesses a contribution that depends on the ratio $\frac{\av{\delta\theta\,\delta\phi}}{\aD\HD}$. Thus, although this correction is second order in perturbations and consequently its amplitude is expected to be small, such correction can grow as the universe expands and eventually it may 
take over the evolution of $\av{\phi}$.

Let us now assume that the previously discussed conditions are fulfilled and consider a homogeneous quintessence field of the form $\phi=\phi(t)=\av{\phi}$.  For this scenario,
expressions (\ref{p_Q})-(\ref{pi_Q}) become
\begin{eqnarray}
\label{p_Q_homogeneous}
p\,&=&\,p_m+\frac{1}{2}(\partial_t\phi)^2-V(\phi)\,,
\\
\label{q_Q_homogeneous}
q_{\mu}&=&0\,,
\\
\label{pi_Q_homogeneous}
\pi^{\mu\nu}&=& g^{\mu\nu}\left[\frac{1}{2}(\partial_t\phi)^2-V(\phi)\right]\,,
\end{eqnarray}
and accordingly $\pi^{\mu\nu}\sigma_{\mu\nu}=0$ and the expression (\ref{jota_definition}) is identically zero. Then, the integrability condition (\ref{generalized_continuity_averaged}) reduces to
\begin{eqnarray}
\frac{1}{2 a_{\mc D}^6}\left[\partial_{t} \left(a_{\mc D}^6 \langle Q \rangle_{\mc D} \right)+a_{\mc D}^4\partial_{t}\left(a_{\mc D}^2 \langle \mathcal{R} \rangle_{\mc D} \right)
\right]\,=\, 
- \langle \theta \,\delta p_m \rangle_{\mc D}\,,
%
\label{extra_terms_Itegrability_Q_homogenous}
\end{eqnarray}
where we have used that the only contribution to the r.h.s. of the previous equation corresponds to the perturbation in the matter pressure term (which vanishes for dust matter) according to  (\ref{p_Q_homogeneous}) under the aforementioned assumption of the homogeneous scalar field. Thus, the 
previous relation proves how the standard integrability condition  is recovered in homogeneous quintessence scenarios whenever
$\delta p_m$ is negligible.
In conclusion, for quintessence models we have seen that thanks to the minimal coupling between gravity and the scalar fields, homogenous scalar fields
do not contribute to the averaged equations. This will no longer be true for theories where the scalar field couples non-minimally, as we show in the next section. However, it is important to keep in mind that homogenous fields are consistent only under the assumptions discussed above.

\subsection{Brans-Dicke theories}
\label{Subsection_BD}
The action for these theories can be written as \cite{JFBD}
\begin{eqnarray}
S\,=\,\frac{1}{2}\int{\rm d}^4x \sqrt{-g}\left[\phi R-\frac{\omega_0}{\phi}\partial_\mu\phi\partial^\mu\phi\right]
+S_M[g_{\mu\nu};\psi],
\label{actionJF}
\end{eqnarray}
where $S_M$ represents the action corresponding to the matter fields $\psi$ and $\omega_0$ is a constant. 
Note that one of the main differences in \eqref{actionJF} with regards to the GR counterpart lies in the fact that the gravitational constant is 
in fact non-constant but depends upon the scalar field $\phi$. The latter contributes to the Lagrangian density with its own kinetic term. In addition, it can be shown that the evolution of the scalar field has as a source term coming from the trace of the matter stress-energy tensor. Thus, the scalar field depends on the mass distribution and consequently the gravitational {\it constant} also does  \cite{Brans:2005ra}.
In the Jordan frame, the modified field equations for Brans-Dicke theories can be written as
\be
\label{Delta}
G_{\mu\nu} =  8\pi G \left( \frac{T_{\mu\nu}^{(m)}}{\phi}+ T_{\mu\nu}^{(\phi)} \right) \,,
\ee
where 
\be
T_{\mu\nu}^{(\phi)} =
\frac{\omega_0}{\phi^2}\left(\nabla_{\mu}\phi\nabla_{\nu}\phi
-\frac{1}{2}g_{\mu\nu}\nabla^{\sigma}\phi\nabla_{\sigma}\phi\right)
+ \frac{\nabla_{\mu}\nabla_{\nu}
\phi-g_{\mu\nu}\Box
\phi  }{\phi}
\; ,
\ee
and the  equation of motion for $\phi$ can be written as
\begin{eqnarray}
\label{Motion_phi}
\Box\phi\,=\,-\frac{\rho_m-3p_{m}}{2\omega_{0}+3}\,.
\end{eqnarray}
For those theories one can calculate $\rho$, $p$, $q^{\mu}$ and $\pi^{\mu\nu}$ as follows
\begin{eqnarray}
\label{mu_BD}
\rho\,&\equiv\,&	T_{\mu\nu}u^{\mu}u^{\nu}\,=\,\frac{\rho_m}{\phi}+\frac{\omega_0}{\phi^2}\left[\dot{\phi}^2+\frac{1}{2}\left(\partial \phi \right)^2\right]+\frac{1}{\phi}\left(\ddot{\phi}+\Box\phi\right)
\,\\
%
\label{p_BD}
p\,&\equiv&\,\frac{1}{3}T_{\mu\nu}h^{\mu\nu}\,=\,\frac{p_m}{\phi}+\frac{\omega_0}{\phi^2}\left[h^{\mu\nu}\hat{\nabla}_{\mu}\phi \hat{\nabla}_{\nu}\phi     -\frac{3}{2}\left(\partial \phi \right)^2 \right]+\frac{1}{3\phi}\left(-2\Box \phi + \ddot{\phi} \right)\,\\
%
\label{q_BD}
q_{\mu}\,&\equiv&\,-T_{\alpha\beta}h^{\alpha}_{\mu}u^{\beta}\,=\,-\frac{\omega_0}{\phi^2}\left[\dot{\phi} \hat{\nabla}_{\mu}\phi  \right]-\frac{1}{\phi}h^{\alpha}_{\;\mu}u^{\beta}\,\nabla_{\beta}\left(h_{\alpha}^{\;\gamma}\nabla_{\gamma}\phi   \right)\,\\
\label{pi_BD}
\pi_{\mu\nu}\,&\equiv&\, T_{\alpha\beta}h^{\alpha}_{\;<\mu}h^{\beta}_{\;\nu>}    \,=\,\hat{T}_{\mu\nu}-p h_{\mu\nu}\;\;\;,\;\;\;  \hat{T}_{\mu\nu}\,\equiv\,h^{\alpha}_{(\mu}h^{\beta}_{\nu)}T_{\alpha\beta}\,.
\end{eqnarray}
Analogously to the procedure sketched in the previous section, let us first decompose the scalar field equation (\ref{Motion_phi}) yielding
\be
\ddot{\phi}+\theta\dot{\phi}-h^{\alpha\beta}\n_\alpha\n_\beta\phi -\frac{1}{2\omega_0-3}\left(\rho_m - 3p_m \right)\,=\,0\,
\ee
By averaging this equation and making repeated use of the commutation relation, we obtain 
\be
\partial_{tt}\av{\phi}+\av{\theta}\partial_t\av{\phi}=\frac{1}{2\omega_0-3}\av{\rho_m-3p_m}+\partial_t\av{\delta\theta\,\delta\phi}+\av{\theta}\av{\delta\theta\,\delta\phi}+\av{h^{\alpha\beta}\n_\alpha\n_\beta\phi}.
\label{BD_averaged}
\ee
This equation is essentially the same as we obtained for the quintessence case with the additional matter-dependent term. However, since this extra term in the local equations determined by the matter fluid is simply the trace of its stress-energy tensor, it contributes a linear term in the corresponding thermodynamic quantities (i.e., $\rho_m$ and $p_m$) and, consequently, the consistency conditions that one needs to consider for a homogeneous field are the same as for quintessence as well as the correction introduced by backreaction effects.

As a first step in understanding the effect of Brans-Dicke fields when subjected to
 averaging, let us consider a homogeneous field $\phi=\phi(t)$ analogously as we did in \ref{Subsection_Quintessence}. In this scenario, expressions 
(\ref{p_BD}) and (\ref{q_BD}) become
\begin{eqnarray}
\label{p_Brans-Dicke_homogeneous}
p\,&=&\,\frac{p_m}{\phi}+\frac{\omega_0}{2}\frac{\dot{\phi}^2}{\phi^2}+\frac{1}{3\phi}\left( \frac{2\rho_m}{2\omega_0+3} +\ddot{\phi}\right)\,,
\\
\label{q_Brans-Dicke_homogeneous}
q_{\mu}\,&=&\,0\,,
\end{eqnarray}
and $\pi^{\mu\nu}$ contracted with $\sigma_{\mu\nu}$ leads to
\begin{eqnarray}
\label{pi_Brans-Dicke_homogeneous}
\pi^{\mu\nu}\sigma_{\mu\nu}\,=\,-\frac{1}{2\phi}\partial_{t}\phi \,\sigma^{ij} \partial_{t} g_{ij}
\end{eqnarray}
%
%
Therefore 
the first term on the r.h.s. of  (\ref{generalized_continuity_averaged}) becomes
\begin{eqnarray}
\langle \theta\, p \rangle_{\mc D} - \langle \theta \rangle_{\mc D}  \langle p \rangle_{\mc D}  \,&=&\,-\frac{2}{3(2\omega_0+3)\phi}  \langle \theta \rangle_{\mc D} \langle \rho_m \rangle_{\mc D}+
 \left \langle \left(\frac{2\rho_m\,\theta}{3(2\omega_0+3)\phi}\right) \right\rangle_{\mc D}\nonumber\\
 &-& \frac{1}{\phi}\left(    \langle \theta p_m \rangle_{\mc D}    -  \langle \theta \rangle_{\mc D}  \langle  p_m \rangle_{\mc D}\right)\nonumber\\
 &=& \frac{-1}{3\phi(2\omega_0+3)}\Big[ \left[\partial_{t},\; \langle  \rangle_{\mc D} \right]\left(2\rho_m+3\left(2\omega_0+3\right)\,p_m \right)\Big]\,.
 %
\label{extra_terms_Itegrability_BD_homogenous}
\end{eqnarray}
Departing from the last result and (\ref{pi_Brans-Dicke_homogeneous}), the integrability condition  (\ref{generalized_continuity_averaged}) yields
\begin{eqnarray}
\frac{1}{2 a_{\mc D}^6}\left[\partial_{t} \left(a_{\mc D}^6 \langle Q \rangle_{\mc D} \right)+a_{\mc D}^4\partial_{t}\left(a_{\mc D}^2 \langle \mathcal{R} \rangle_{\mc D} \right)
\right]\,&=&\,\frac{-1}{3\phi(2\omega_0+3)}\Big[ \left[\partial_{t},\; \langle  \rangle_{\mc D} \right]\left(2\rho_m+3\left(2\omega_0+3\right)\,p_m \right)\Big]
\nonumber\\
&-&\frac{1}{2\phi}\partial_{t}\phi \left\langle \sigma^{ij}\partial_t g_{ij}\right\rangle_{\mc D}
\label{Itegrability_BD_homogenous}
\end{eqnarray}
This relation proves how the standard integrability condition is not recovered when a homogeneous scalar field is present provided the matter 
density and pressure, the metric tensor and $\sigma_{ij}$ are 
inhomogeneous. Thus we see how, unlike the results for quintessence, the Brans-Dicke non-minimal coupling between gravity and the scalar field provides significant differences with respect to standard averaging in GR. 

\section{Towards the Einstein frame: conformal transformations}
\label{EinsteinFrame}
Let us consider again the action (\ref{actionJF}) including now a potential term for the scalar field,
\be
S_{BD}=\int {\rm d}^4x\sqrt{-g}\left[\phi R-\frac{\omega_0}{\phi}g^{\mu\nu}\nabla_{\mu}\phi\nabla_{\nu}\phi-V(\phi)+2  \mathcal{L}_m\right]\ .
\label{D1.4}
\ee
where $\omega_0$ is a constant. The action (\ref{D1.4}) represents the gravitational action of non-minimally coupling scalar-tensor theories expressed in the so-called Jordan frame. Thus, the field equations are obtained by varying the action (\ref{D1.4}) with respect to the metric tensor  $g_{\mu\nu}$ and the scalar field $\phi$,
\[
R_{\mu\nu}-\frac{1}{2}g_{\mu\nu}R=\frac{1}{\phi}T_{\mu\nu}^{(m)}+\frac{\omega_0}{\phi^2}\left[\nabla_{\mu}\phi\nabla_{\nu}\phi-\frac{1}{2}g_{\mu\nu}\nabla^{\sigma}\phi\nabla_{\sigma}\phi\right]+\frac{1}{\phi}(\nabla_{\mu}\nabla_{\nu}\phi-g_{\mu\nu}\Box\phi)-\frac{1}{2}g_{\mu\nu}V(\phi)\ ,
\]
\be
(2\omega_0+3)\Box\phi\,=\,T^{(m)}+\phi\frac{dV(\phi)}{d\phi}-2V(\phi)\ .
\label{1.5}
\ee
The action (\ref{D1.4}) can then be rewritten in the so-called Einstein frame by applying the following conformal transformation,
\be
g_{E\mu\nu}=\Omega^2g_{ \mu\nu},  \quad \text{where} \quad \Omega^2=\phi\ ,
\label{D2.1}
\ee
which cancels the non-minimally coupling term of the action (\ref{D1.4}) leading to 
\be
S_{E}=\int {\rm d}^4x\sqrt{-g_E}\left[R_E-\frac{2\omega_0+3}{2\phi^2}\partial_{\mu}\phi\partial^{\mu}\phi-\frac{V(\phi)}{\phi^2} +\frac{2}{\phi^2}\mathcal{L}_{Em}\right]\ .
\label{D2.2}
\ee
where the subscript $_E$  refers to the Einstein frame whereas the matter Lagrangian is given by $\mathcal{L}_{Em}=\mathcal{L}_m\left(\phi,g_{E\mu\nu}\right)$. In order to simplify the action (\ref{D2.2}), the scalar field  can be redefined as $\phi=\e^{\varphi/\sqrt{3+2\omega_0}}$, which yields
\be
S_{E}=\int d^4x\sqrt{-g_E}\left[R_E-\frac{1}{2}\partial_{\mu}\varphi\partial^{\mu}\varphi-U(\varphi) +2\alpha(\varphi)\mathcal{L}_{Em}\right]\ ,
\label{D2.3}
\ee
where $\alpha(\varphi)=\frac{1}{\phi(\varphi)^2}$ and $U(\varphi)=\e^{\varphi/\sqrt{3+2\omega_0}}V(\phi(\varphi))$, whereas the field equations are transformed as
\be
 R_{E\mu\nu}-\frac{1}{2}g_{E\mu\nu}R_E=T_{\mu\nu}\ ,
\label{D2.4}
\ee
\be
\Box\varphi-\frac{dU(\varphi)}{d\varphi}=-2\frac{\delta(\alpha(\varphi)\mathcal{L}_{Em})}{\delta\varphi}\ ,
\label{D2.5}
\ee
where $T_{\mu\nu}=T_{\mu\nu}^{(\varphi)}+\alpha(\varphi)T^{(m)}_{E\mu\nu}$, and
\be
T_{E\mu\nu}^{(m)}=\frac{-2}{\sqrt{-g_E}}\frac{\delta\mathcal{L}_{E\mathrm{m}}}{\delta g_E^{\mu\nu}}
\label{MEMT}
\ee
\be
T_{\mu\nu}^{(\varphi)}=\frac{-2}{\sqrt{-g_E}}\frac{\delta S_{\varphi}}{\delta g_E^{\mu\nu}}=\partial_{\mu}\varphi \partial_{\nu}\varphi-g_{E\mu\nu}\left(\frac{1}{2}\partial_{\sigma}\varphi \partial^{\sigma}\varphi+U(\varphi)\right)
\label{SEMT}
\ee
Then, we can implement the averaging procedure in the framework of the action (\ref{D2.3}) by following the same procedure as in previous sections.

\subsection{Averaging procedure in the Einstein frame}

From now on, we explore the averaging procedure for the action (\ref{D2.3}), so the subscript $_E$ is omitted for clearness. As usual, we assume a set of observers described by a unitary timelike 4-velocity vector $u^{\mu}$. Let us consider the energy constraint, the momentum constraint and the Raychaudhuri equations (\ref{eqn_averaged_1}-\ref{eqn_averaged_3})
and define the total energy-momentum tensor as $T_{\mu\nu}=T_{\mu\nu}^{(\varphi)}+\alpha(\varphi)T^{(m)}_{E\mu\nu}=\rho u_{\mu}u_{\nu}+ph_{\mu\nu}+2q_{(\mu}u_{\nu)}+\pi_{\mu\nu}$, where the thermodynamical quantities are given by 
\be
\rho=T_{\mu\nu}u^{\mu}u^{\nu}=\frac{1}{2}\dot{\varphi}^2+\frac{1}{2}h^{\mu\nu}\hat{\nabla}_{\mu}\varphi\hat{\nabla}_{\nu}\varphi+U(\varphi)+\alpha(\varphi)\rho_m\ ,
\label{A3}
\ee
\be
p=\frac{1}{3}T_{\mu\nu}h^{\mu\nu}=\frac{1}{2}\dot{\varphi}^2-\frac{1}{6}h^{\mu\nu}\hat{\nabla}_{\mu}\varphi\hat{\nabla}_{\nu}\varphi-U(\varphi)+\alpha(\varphi)p_m\ ,
\label{A4}
\ee
\be
q_{\lambda}=-T_{\mu\nu}h^{\mu}_{\lambda}u^{\nu}=-\dot{\varphi}\varphi_{,\lambda}\ ,
\label{A5}
\ee
\be
\pi_{\mu\nu}=T_{cd}h^{c}_{\langle\mu}h^{d}_{\nu\rangle}=T_{cd}h^{c}_{(\mu}h^{d}_{\nu)}-ph_{\mu\nu}\ 
\label{A6}
\ee
whereas the zero component of the continuity equation yields 
\be
\dot{\rho}+\theta(\rho+p)+\nabla_{\mu}q^{\mu}+\pi^{\mu\nu}\sigma_{\mu\nu}+a^{\mu}q_{\mu}=0.
\ee 
Then, in this frame the Buchert's equations are obtained by averaging the equations (\ref{eqn_averaged_1})-(\ref{eqn_averaged_3}), 
where in this case the averaged energy density and pressure (\ref{A3}-\ref{A4}) become
\be
\langle\rho\rangle_{\mc D}=\frac{1}{2}\langle\dot{\varphi}^2\rangle_{\mc D}+\frac{1}{2}\langle h^{\alpha\beta}\hat{\nabla}_{\alpha}\varphi\hat{\nabla}_{\beta}\varphi \rangle_{\mc D}+\langle U(\varphi)\rangle+\langle\alpha(\varphi)\rho_m\rangle_{\mc D}\ .
\label{A3a}
\ee
\be
\langle p\rangle_{\mc D}=\frac{1}{2}\langle\dot{\varphi}^2\rangle_{\mc D}-\frac{1}{6}\langle h^{\alpha\beta}\hat{\nabla}_{\alpha}\varphi\hat{\nabla}_{\beta}\varphi \rangle_{\mc D}-\langle U(\varphi)\rangle_{\mc D}+\langle\alpha(\varphi)p_m\rangle_{\mc D}\ .
\label{A4a}
\ee
Whereas the continuity equation yields
\be
\partial_t\av{\rho}+\langle\rho\rangle_{\mc D}\langle\theta\rangle_{\mc D}=-\langle p\theta\rangle_{\mc D}-\langle\pi_{\mu\nu}\sigma^{\mu\nu}\rangle_{\mc D}-\langle a^{\mu}q_{\mu}\rangle_{\mc D}
\label{A10}
\ee
%
%
%
Assuming $a^{\mu}=0$ and considering (\ref{A3})-(\ref{A6}) one recovers the integrability condition 
(\ref{generalized_continuity_averaged}).
Let us now explore the case of a homogeneous scalar field $\varphi({\bf x},t)=\varphi(t)$ coupled to a dust fluid, $p_m=0$.  In such a case, $q_{\lambda}=0$, 
and the  integrability condition 
(\ref{generalized_continuity_averaged}) leads to
\be
%
\frac{1}{2a_{\mc D}^6}\Big[\partial_{t} \left(a_D^6 \av{Q}  \right)+a_D^4\partial_{t}\left(a_D^2 \av{\mc R}\right)
\Big]\,
\,=\,\frac{1}{2}\Big[ \partial_{t},\,\av{}\Big] \left(\frac{1}{2}\dot{\varphi}^2-U(\varphi)\right)
\label{A12}
\ee
which turns out to the usual integrability condition when a pressureless fluid is considered. Thus,  
the integrability condition might be recovered despite the presence of a strong coupling between pressure-free matter and the scalar field, 

\subsection{Averaging in different frames}
%
%
%
%

In section \ref{Subsection_BD} we have discussed the averaging for the Brans-Dicke class of scalar-tensor theories 
and actions with non-minimally couplings among matter and a scalar field as given by Eq. (\ref{D2.3}). As it is well-known, a sub-class of Brans-Dicke-like theories without kinetic term but non-vanishing scalar field potential can be mapped into $f(R)$ theories where the role of the scalar field is essentially played by $f_R\equiv\partial f/\partial R$. In such theories, one can again describe the modified Einstein equations by means of an effective geometrical fluid with stress-energy tensor
\begin{eqnarray}
T^{(f)}_{\mu\nu}\,\equiv\,\frac{1-f_R}{f_R}T^{(m)}_{\mu\nu}+
\frac{1}{f_{R}}\left[
(\nabla_{\mu}\nabla_{\nu}-g_{\mu\nu}\Box)
f_{R}-\frac{1}{2}\left(f(R)-R\,f_R\right)g_{\mu\nu}\right]
\label{eff_energy_momentum_tensor}
\end{eqnarray}
and apply our general formulae to obtain the modified averaged equations together with the integrability condition. In particular, we can obtain the expression for the modified integrability condition as given in (\ref{generalized_continuity_averaged}). A crucial feature that renders calculations more complicated in this scenario lies in the fact that the effective fluid depends on second derivatives of the Ricci curvature $R$ so that the the r.h.s of the integrability condition will depend on second derivatives of $R$. Thus the integrability condition will no longer be a first order differential equation relating the kinematical backreaction and the averaged curvature. Although this does not pose any conceptual problem, it introduces technical difficulties.
In order to get around this problem appearing in several extended gravity theories, one may try to perform calculations in the Einstein frame, where the action (\ref{D1.4}) is mapped into the familiar Einstein-Hilbert action, but with the presence of non-trivial couplings in the matter sector given by Eq. (\ref{D2.3}) and this should facilitate the analysis, as shown previously. However, one must be aware that in order to do so, a conformal transformation needs to be performed.  
The conformal transformation implies several effects on 
the averaging procedure, since averaging and frame changing constitute non-commutative operations. 
Furthermore, let recall that the spatial average of a scalar quantity is given by
(\ref{average_definition}).
Then, by applying the conformal transformation (\ref{D2.1}) $g_{E ij}=\phi g_{ij}$, 
the determinant of the spatial metric is transformed as $g_E^{(3)}=\Omega^6 h$. The 
comparison of Riemannian volume elements in both frames thus leads to
\be
\diff\Sigma=
\sqrt{h}\,\diff^3x=
\Omega^3(t,{\bf x})\sqrt{g_E^{(3)}}\,\diff^3x
=\Omega^3(t,{\bf x})\,\diff\Sigma_E.
\label{AB4}
\ee
Then, the Riemannian volumes of the domains do not have a straightforward correspondence. 
In fact, it is well known that the Einstein and Jordan frames do not lead to the same equations, although solutions obtained in each frame 
can be easily related by the conformal transformation (\ref{D2.1}) and a particular redefinition of the coordinates
%
%
(see for instance \cite{Faraoni:2006fx} and references therein), leading to a correspondence between both frames that might become very useful while analysing the properties of a particular theory. In addition, any conformally invariant physical quantity remains the same in both frames. Nevertheless, when applying the averaging procedure, it is straightforward to note that even for a conformally invariant quantity, i.e., $\mc O\sqrt{h}={\mc O}_E\sqrt{g_E^{(3)}}$ invariant under conformal transformations, the average would not remain invariant since the domain volume does not have a clear correspondence between frames (\ref{AB4}). Indeed, even if the underlying theory is conformally invariant, the averaging procedure will introduce a characteristic scale, namely the size of the domains, that will explicitly break such an invariance.
However, note that the domain of integration, where the average is performed, would rescale also through the conformal transformation, but even by redefining the domain, the integral is evaluated along a different path since the conformal transformation depends on the coordinates. 
Moreover, the domain of integration in the Einstein frame might not even be well defined. Note that the domain of integration can be defined using 
two hypersurfaces of interest: either with respect to the gravitational frame, which refers to a set of comoving observers defined at every point of the spacetime, or with respect to the rest frame of the fluid \cite{Umeh:2010pr}. In the latter case, while referring to the rest frame of the matter content, one should be aware that in the Einstein frame there exists a fifth force mediated by the scalar field that will make the matter trajectories non-geodesic \cite{Olive,Gilles}, and $\nabla_{E\mu}T_{E}^{\mu\nu(m)}\neq 0$. This is natural since in the Einstein frame there is an exchange of energy between the matter and the scalar field. This fact introduces difficulties 
when defining the domain of integration with respect to the matter content. The geodesic character can be recovered if congruences are defined with respect to the flux
at rest with respect the total energy-momentum tensor. 
Furthermore, if one departs from the Jordan frame, where the acceleration of a set of observers  
$a_{\mu}=u^{\nu}\nabla_{\nu}u_{\mu}$ is assumed to vanish, i.e., it constitutes a geodesic congruence, and then, performs a conformal transformation (\ref{D2.1}), the 4-acceleration in the Einstein frame becomes
\be
a_{E\mu}=u^{\nu}_E\nabla_{E\nu}u_{E\mu}=\Omega^{-1}\left(\partial_{\mu}\Omega+u_{\mu}u^{\lambda}\partial_{\lambda}\Omega\right)\ ,
\label{Ab7}
\ee
where $u^{\mu}_{E}=\Omega^{-1}u^{\mu}$ with $u^{\mu}u_{\mu}=\Omega^{-2}g_{E\mu\nu}u^{\mu}u^{\nu}=g_{E\mu\nu}u^{\mu}_Eu^{\nu}_E=-1$. Unless the scalar field
$\phi\equiv\Omega^2$ is exactly homogeneous, the acceleration (\ref{Ab7}) in the Einstein frame will not vanish and this term will eventually introduce additional corrections in the corresponding results for the local equations.

To summarise the above discussion, we can conclude that the relation between both frames can not be clearly established when dealing with the averaging over the physical quantities. Although in principle one could apply the averaging procedure in any frame, computations seem to be more straightforward and technically less challenging in the Jordan frame where the gravitational sector is described by a scalar-tensor theory and matter is minimally coupled to gravity. However, when minimal couplings are assumed in one particular frame, non-minimal couplings will emerge when applying a conformal transformation to the original frame,  so that a set of observers become accelerated in the new frame. Moreover, as pointed out in Refs.~\cite{Sokolowski:1989ee,Magnano:1993bd}, also the physical meaning of each frame can not be easily established. An  analysis of the ground state and the positivity of the energy in both frames is an important tool to determine whether the frames, and in particular the Jordan one, are well defined. In addition, non-minimal couplings between matter and the scalar field may be avoided in the Einstein frame by  transforming them to the Jordan frame. In this case observers would follow geodesics in the Einstein frame and consequently avoid the appearance of a non-zero acceleration (see Ref.~\cite{Magnano:1993bd}). Nevertheless, in the latter case, the observers in the Jordan frame will not follow geodesics, since a fifth force is induced by the coupling between matter and the scalar field. Hence, the choice of the most suitable frame and their physical significance - an old problem that has been widely studied in the literature \cite{Faraoni_Review} - would depend on the underlying theory as well as on other aspects that affect the physical viability of each frame. In our particular discussion, we have pointed out to the problems and differences emerging when the frame is conformally transformed together with the averaging procedure, and the difficulties occurring when the calculations in both frames try to be related.
%
%
%
\section{Discussion and Conclusions}
\label{conclusions}
%
In this paper we have introduced a generalised Buchert backreaction procedure in a scenario with two non-interacting fluids. The set-up is aimed to account for non-standard cosmologies where, in addition to the usual matter, there is an additional contribution to the total stress-energy tensor sourcing Einstein's equations. This additional contribution might correspond to a dark energy component or an effective description of a modified gravity theory. We started by deriving the relevant local equations in the usual ADM decomposition. In order to simplify our calculations, we chose the rest frame of the matter component, which remains geodesic at all times.

Equipped with the local equations, we extended the standard Buchert equations and obtained a generalisation for general fluids with
pressure, shear, anisotropic stress and momentum flux. The combination of the averaged equations allowed us to obtain generalised integrability condition as given in expression (\ref{generalized_continuity_averaged}). We have shown how this expression is modified with respect to the usual GR case with matter. This correction is manifestly second order in the perturbations. Moreover, the non-zero character of the right-hand side in the aforementioned equation may cause difficulties in order to close the averaged system of equations when, for instance, results from perturbative backreaction are to be explored by calculating second order terms around a given background. We also argued that this new term can be crucial in modified cosmologies with dark energy or alternative gravity theories, because it will modify the kinematical backreaction evolution. We have illustrated this for a simple case in which the spatial scalar curvature averages to that of a FLRW metric. Thus, we have shown how, even in cases where backreaction is negligible for GR with matter, it might give rise to non-negligible effects for modified cosmologies and this could play a crucial role in testing such alternative scenarios. 

We applied our general formalism and the generalised Buchert equations to three classes of extended scenarios, which can be considered as natural extensions of GR: a dark energy model described by a perfect fluid, a quintessence field and, finally, Brans-Dicke theories. In the dark energy model described by a perfect fluid we assumed a barotropic equation of state and showed how it can modify the equation of state for averaged energy density and pressure. Moreover, for constant equation of state parameter and assuming a power law evolution for the backreaction source term in both the continuity equation and integrability condition, we obtained the solution for the average energy density and the kinematical backreaction. For this very simple case, we determined the conditions under which the backreaction effects are important and even modify the usual homogeneous evolution. Although we have focused on the simplest case of a dark energy fluid with constant equation of state and adiabatic perturbations, it would be interesting to extend our analysis to more general frameworks when dark energy can have non-adiabatic perturbations or be described by an imperfect fluid. In this respect, the effective field theory for cosmological fluids developed for one single perfect fluid \cite{Ballesteros:2012kv} or a multicomponent scenario \cite{Ballesteros:2013nwa} might be appealing.

For extensions based on a single scalar field (quintessence and Brans-Dicke) we obtained the averaged version of the field equations. In particular, we obtained the evolution equation for the averaged scalar field and shown how backreaction can effect such evolution. From these equations, we obtained the conditions under which it is consistent to consider a purely homogeneous field such that its evolution coincides with that of its average, i.e., $\phi(t)=\av{\phi}$. We also computed the evolution of $\av{\phi}$ by assuming power-law evolutions for the backreaction source terms, very much like in the perfect fluid model. With these solutions we again determined conditions for the backreaction effects to be relevant. We then assumed that the scalar field is purely homogeneous and obtained how, even though being homogeneous, it can give non-trivial contributions to the averaged Buchert equations and the integrability condition. For the quintessence mode, being minimally coupled, no effects arise, but for the non-minimally coupled field present in the Brans-Dicke models, the homogeneous scalar field can give non-trivial contribution. The analysis presented in this communication could be further extended to include more general single field models and to consider the more general case of inhomogeneous fields.

The averaging procedure has also been implemented when the matter content is coupled to a scalar field, which occurs in the so-called Einstein frame when a conformal transformation is applied to Brans-Dicke-like theories. Furthermore, a brief discussion dealing with the comparison of both frames was provided. In this sense, the mapping to the Einstein frame seems to be problematic when analysing the average of extended gravity theories despite the fact that the equations may look simpler in the Einstein frame. The point is that non-minimal couplings between matter and the scalar field are induced when applying a conformal transformation to the usual Brans-Dicke-like action. After the aforementioned transformation, geodesic observers become accelerated.  Moreover, while working with non-minimal couplings between matter sources and the scalar field,
one should be aware of the definition of the integration domain, since the matter content will not follow geodesics in the conformal frame, and the total flow over all fields has to be considered in order to get an appropriate definition of the integration domain. In addition, 
the relation between averaged quantities in the two frames remains unclear since the domain volumes in different frames don't have a clear correspondence either.  On the other hand, besides the mathematical tool that the conformal transformation may provide,  the physical meaning of both frames has to be analysed carefully, where the positivity of the energy and the behaviour of the ground state may play a crucial role to discriminate between frames, as pointed out in Refs~\cite{Sokolowski:1989ee,Magnano:1993bd}. 

The theoretical tools provided in this communication are easily extendible to other alternative gravity theories
as well as scenarios combining  gravitational theories beyond General Relativity with standard fluids different from dust. 
The range of different techniques able to determine the evolution of both background and perturbations in extended gravity theories also permits one to perform perturbative averaging. This way - and together with the equations presented in this communication - one can estimate the size of the backreaction effect and the way in which the fluctuations become of the order of the mean, leading to situations where the assumed background will no longer correctly describe the averages. In this way, fluctuations may affect the background which needs to be confronted with 
with large scale observables, such as the luminosity distances and Baryon Acoustic Oscillations.

%
\vspace{0.7cm}
{\bf Acknowledgments:} 
We would like to thank M. Seikel for useful comments to start this investigation. 
We are also indebted to C. Clarkson and J. P. Uzan for useful discussions. 
J.B.J. is supported by the Wallonia-Brussels Federation grant ARC No.~11/15-040. 
A.d.l.C.D. acknowledges financial support from MINECO (Spain) project FPA2011-27853-C02-01. 
A.d.l.C.D. thanks the financial support in 2013 from a Marie Curie - Beatriu de Pin\'os contract BP-B00195 Generalitat de Catalunya and a ACGC fellowship University of Cape Town.
 J.B.J. and A.d.l.C.D. thank projects FIS2011-23000 and Consolider-Ingenio MULTIDARK CSD2009-00064 for financial support.
P.K.S.D. thanks the NRF for financial support. 
D. S.-G. acknowledges the support from the University of the Basque Country, Project Consolider CPAN Bo. CSD2007-00042, the URC financial support from the University of Cape Town (South Africa) and MINECO (Spain) project FIS2010-15640. J.B.J. wishes to thank the Department of Mathematics and Applied Mathematics and ACGC, University of Cape Town for their warm hospitality.%


\smallskip


\medskip
\begin{thebibliography}{9}
%
\bibitem{Riess}
  A. G. Riess et al. [Supernova Search Team Collaboration], Astron. J. 116, 1009, (1998);
  S. Perlmutter et al.[Supernova Cosmology Project Collaboration], Astro- phys. J. 517, 565, (1999).

\bibitem{Rasanen_Feb_11} 
  S.~Rasanen,   
  Class.\ Quant.\ Grav.\  {\bf 28} (2011) 164008. 

  
  \bibitem{Refs_GR_wrong}
  P.~J.~E.~Peebles,   
  astro-ph/0410284;
%
  M.~Vonlanthen, S.~Rasanen and R.~Durrer,   
  JCAP {\bf 1008} (2010) 023; 
  %
  L.M. Krauss and B. Chaboyer, 
Constraints on Cosmology, Science {\bf 299} (2003) 65;
  A.~G.~Riess {\it et al.},  
  Astrophys.\ J.\  {\bf 730} (2011) 119    [Erratum-ibid.\  {\bf 732} (2011) 129]; 
  Astrophys.\ J.\  {\bf 699} (2009) 539 
    
\bibitem{Copeland}
  E. J. Copeland, M. Sami and S.Tsujikawa, {\it Int. J. Mod. Phys.} {\bf D15}, 1753 (2006).

\bibitem{Quintessence}
%
R.R. Caldwell, R. Dave, and P.J. Steinhardt, {\it Phys. Rev. Lett.} {\bf 80}, 1582 (1998);
  E.~Elizalde, S.~'i.~Nojiri, S.~D.~Odintsov, D.~Saez-Gomez and V.~Faraoni, 
  Phys.\ Rev.\ D {\bf 77}, 106005 (2008), [arXiv:0803.1311 [hep-th]];
  %
  J.~Beltr\'an~Jim\'enez, P.~Santos and D.~F.~Mota,
  Phys.\ Lett.\ B {\bf 723} (2013) 7
  [arXiv:1212.5266 [astro-ph.CO]].
  
  \bibitem{K-essence}
  T.~Chiba, T.~Okabe and M.~Yamaguchi,
  Phys.\ Rev.\ D {\bf 62} (2000) 023511
  [astro-ph/9912463].\\
  C.~Armendariz-Picon, V.~F.~Mukhanov and P.~J.~Steinhardt,
  Phys.\ Rev.\ Lett.\  {\bf 85} (2000) 4438
  [astro-ph/0004134].\\
  C.~Armendariz-Picon, V.~F.~Mukhanov and P.~J.~Steinhardt,
  Phys.\ Rev.\ D {\bf 63} (2001) 103510
  [astro-ph/0006373].

\bibitem{Lovelock}
  C.~Lanczos, Z. Phys. {\bf 73}, 147, (1932); Annals Math. {\bf 39}, 842, (1938);
  D.~Lovelock, J.\ Math.\ Phys.\  {\bf 12}, 498 (1971).

\bibitem{GB}
  G. Cognola, E. Elizade, S. Nojiri, S. D. Odintsov and S. Zerbini, Phys. Rev. D \textbf{73} 084007 (2006);  [arxiv:hep-th/0601008].
  S.~Nojiri, S.~D.~Odintsov, Phys. Lett. B \textbf{631} 1 (2005);  [arxiv:hep-th/0508049];
  Phys. Rev. D \textbf{68}, 123512 (2003); [hep-th/0307288];
  E.~Elizalde, R.~Myrzakulov, V.~V.~Obukhov and D.~S\'aez-G\'omez, Class. Quant. Grav. \textbf{27}  095007 (2010);   [arXiv:1001.3636 [gr-qc]].
  R.~Myrzakulov, D.~S\'aez-G\'omez and A.~Tureanu, Gen.\ Rel.\ Grav.\ \ {\bf 43} 1671 (2011); [arXiv:1009.0902 [gr-qc]].
  A. de la Cruz-Dombriz and D. S\'aez-G\'omez, Class. Quantum Grav. {\bf 29}  245014, (2012), arXiv:1112.4481 [gr-qc].

\bibitem{JFBD} 
  P. Jordan, Schwerkaft und Weltall (Vieweg, Braunschweig, 1955); M. Fierz, Helv. Phys. Acta 29, 128 (1956); C. Brans and R.H. Dicke, Phys. Rev. 124, 925 (1961).

\bibitem{Brans:2005ra}
  C.~H.~Brans, The Roots of scalar-tensor theory: An Approximate history, [gr-qc/0506063].
  
 \bibitem{ST}
  C. H. Brans, Phys. Rev., {\bf 125(6)} 2194 (1962);
  J. Garc\'ia-Bellido, A. Linde, and D. Linde, Phys. Rev. D, {\bf 50} 730 (1994);
  J.~A.~R.~Cembranos  {\it et al.}, JCAP {\bf 0907}, 025 (2009); 
  T.~Biswas {\it et al.}, Phys.\ Rev.\ Lett.\  {\bf 104}, 021601 (2010); 
  JHEP {\bf 1010}, 048 (2010); 
  Phys.\ Rev.\  D {\bf 82}, 085028 (2010). 
 
\bibitem{generalST}
  C.~Deffayet, O.~Pujolas, I.~Sawicki and A.~Vikman,
  JCAP {\bf 1010} (2010) 026;
  C.~de Rham and L.~Heisenberg,
  Phys.\ Rev.\ D {\bf 84} (2011) 043503;
  J.~Beltr\'an~Jim\'enez, E.~Dio and R.~Durrer,
  JHEP {\bf 1304} (2013) 030.
 
 \bibitem{VT}
  L.~H.~Ford, Phys.\ Rev.\ D {\bf 40} (1989) 967.
  J.~Beltr\'an Jim\'enez and A.~L.~Maroto, Phys.\ Rev.\ D {\bf 78} (2008) 063005; JCAP {\bf 0903} (2009) 016;
  Phys.\ Rev.\ D {\bf 80} (2009) 063512;
  T.~Koivisto and D.~F.~Mota, JCAP {\bf 0808}, 021 (2008); 
  J.~A.~R.~Cembranos  {\it et al.}, Phys.\ Rev.\ D {\bf 86}, 021301 (2012); 
  arXiv:1212.3201 [astro-ph.CO]. 
  J.~Beltr\'an~Jimenez, A.~L.~Delvas Froes and D.~F.~Mota,
  Phys.\ Lett.\ B {\bf 725}, 212 (2013)
  [arXiv:1212.1923 [astro-ph.CO]].
  J.~Beltr\'an Jim\'enez, R.~Durrer, L.~Heisenberg and M.~Thorsrud,
  JCAP {\bf 1310}, 064 (2013)
  [arXiv:1308.1867 [hep-th]].

 \bibitem{XD}
  J.~Alcaraz {\it et al.}, Phys. Rev.{\bf D67}, 075010 (2003); 
  P. Achard {\it et al.}, Phys. Lett. {\bf B597}, 145 (2004); 
  J.~A.~R.~Cembranos, A.~Dobado and A.~L.~Maroto,  Phys.\ Rev.\ Lett.\  {\bf 90}, 241301 (2003); 
  Phys.\ Rev.\ D {\bf 68}, 103505 (2003); 
  AIP Conf.Proc. {\bf 670}, 235 (2003); 
  Int. J. Mod. Phys. {\bf D13}, 2275 (2004); 
  Phys. Rev. {\bf D70}, 096001 (2004); 
  Phys.\ Rev.\ D {\bf 73}, 035008 (2006); 
  Phys.\ Rev.\ D {\bf 73}, 057303 (2006); 
  J.\ Phys.\ A  {\bf 40}, 6631 (2007); 
  J.~A.~R.~Cembranos, R.~L.~Delgado and A.~Dobado, arXiv:1306.4900 [hep-ph].  

\bibitem{sugra}
  D.~Z.~Freedman, P.~van Nieuwenhuizen and S.~Ferrara, Phys.\ Rev.\ D {\bf 13}, 3214 (1976); 
  S.~Deser and B.~Zumino, Phys.\ Lett.\ B {\bf 62}, 335 (1976);  
  E.~Cremmer, B.~Julia and J.~Scherk, Phys.\ Lett.\ B {\bf 76}, 409 (1978); 
  L.~J.~Hall, J.~D.~Lykken and S.~Weinberg, Phys.\ Rev.\ D {\bf 27}, 2359 (1983); 
  N.~Ohta, Prog.\ Theor.\ Phys.\  {\bf 70}, 542 (1983); 
  L.~Alvarez-Gaume, J.~Polchinski and M.~B.~Wise,  Nucl.\ Phys.\ B {\bf 221}, 495 (1983); 
  H.~P.~Nilles, Phys.\ Rept.\  {\bf 110}, 1 (1984). 
  J.~A.~R.~Cembranos, J.~L.~Feng, A.~Rajaraman and F.~Takayama, Phys.\ Rev.\ Lett.\  {\bf 95}, 181301 (2005); 
  AIP Conf.\ Proc.\  {\bf 903}, 591 (2007). 
  J.~A.~R.~Cembranos, J.~L.~Feng and L.~E.~Strigari, Phys.\ Rev.\ Lett.\  {\bf 99}, 191301 (2007); 
  Phys.\ Rev.\  D {\bf 75}, 036004 (2007); 
  M.~R.~Garousi, arXiv:1210.4379 [hep-th].

 \bibitem{disformal}
  G.~W.~Horndeski, Int.\ J.\ Theor.\ Phys.\  {\bf 10}, 363 (1974);  
  J.~D.~Bekenstein, Phys.\ Rev.\ D {\bf 48}, 3641 (1993); 
  J.~A.~R.~Cembranos {\it et al.}, Phys. Rev. {\bf D65} 026005 (2002); 
  JCAP {\bf 0810}, 039 (2008); 
  Phys.\ Rev.\  D {\bf 83}, 083507 (2011); 
  Phys.\ Rev.\ D {\bf 84}, 083522 (2011); 
  Phys.\ Rev.\ D {\bf 85}, 043505 (2012); 
  arXiv:1204.0655 [hep-ph]; 
  arXiv:1305.2124 [hep-ph]; 
  JCAP {\bf 1304}, 051 (2013);  
  J.~A.~R.~Cembranos and L.~E.~Strigari, Phys.\ Rev.\  D {\bf 77}, 123519 (2008); 
  M.~Zumalacarregui, T.~S.~Koivisto, D.~F.~Mota and P.~Ruiz-Lapuente, JCAP {\bf 1005}, 038 (2010); 
  T.~S.~Koivisto, D.~F.~Mota and M.~Zumalacarregui,    
  Phys.\ Rev.\ Lett.\  {\bf 109} (2012) 241102; 
  Phys.\ Rev.\ D {\bf 87} (2013) 083010. 

 \bibitem{LV}
  V. A. Kostelecky and S. Samuel, Phys. Rev. D {\bf 39}, 683 (1989).
  D. Colladay and V. A. Kostelecky, Phys. Rev. D {\bf 55}, 6760 (1997);
  J. R. Ellis, N. E. Mavromatos and D. V. Nanopoulos,  Phys. Rev. D {\bf 61}, 027503 (1999);
  J. Alfaro, H. A. Morales-Tecotl and L. F. Urrutia, Phys. Rev. Lett. {\bf 84}, 2318 (2000);
  G. Amelino-Camelia, Nature {\bf 418}, 34 (2002);  Int.\ J.\ Mod.\ Phys.\ D {\bf 11}, 35 (2002)  
   G.~Amelino-Camelia, J.~R.~Ellis, N.~E.~Mavromatos, D.~V.~Nanopoulos and S.~Sarkar,
  Nature {\bf 393}, 763 (1998);
  J. Magueijo and L. Smolin, Phys. Rev. Lett. {\bf 88}, 190403 (2002);
  J.~A.~R.~Cembranos, A.~Rajaraman and F.~Takayama, hep-ph/0512020; 
  Europhys.\ Lett.\  {\bf 82}, 21001 (2008); 
  J.~A.~R.~Cembranos, arXiv:1301.7088 [hep-ph]; %
  S. Ghosh and P. Pal, Phys. Rev. D {\bf 75}, 105021 (2007).



\bibitem{fR}
  T.~P.~Sotiriou, J.\ Phys.\ Conf.\ Ser.\  {\bf 189}, 012039 (2009); 
  J.~A.~R.~Cembranos, Phys.\ Rev.\  D {\bf 73}, 064029 (2006); 
  Phys.\ Rev.\ Lett.\  {\bf 102}, 141301 (2009); 
  S.~Capozziello and M.~De Laurentis, Phys.\ Rept.\  {\bf 509}, 167 (2011); 
  A.~de la Cruz-Dombriz, A.~Dobado and A.~L.~Maroto,  Phys.\ Rev.\  D {\bf 77} (2008) 123515; 
%
 A.~Abebe, M.~Abdelwahab, A.~de la Cruz-Dombriz and P.~K.~S.~Dunsby,   
  Class.\ Quant.\ Grav.\  {\bf 29}, 135011 (2012); 
  S.~'i.~Nojiri and S.~D.~Odintsov, Phys.\ Rept.\  {\bf 505}, 59 (2011); 
  T.~P.~Sotiriou and V.~Faraoni, Rev.\ Mod.\ Phys.\  {\bf 82}, 451 (2010);  
  J. A. R. Cembranos {\it et al.}, JCAP {\bf 1204}, 021 (2012); 
  AIP Conf. Proc. {\bf 1458}  491 (2011);
  A.~de la Cruz-Dombriz and D.~S\'aez-G\'omez, Entropy {\bf 14}, 1717 (2012),  [arXiv:1207.2663 [gr-qc]];
  F.~D.~Albareti  {\it et al.}, JCAP {\bf 1212}, 020 (2012); 
  arXiv:1212.4781 [gr-qc]. 
T.~Clifton, P.~Dunsby, R.~Goswami and A.~M.~Nzioki, 
  Phys.\ Rev.\ D {\bf 87}, no. 6, 063517 (2013)
  [arXiv:1210.0730 [gr-qc]];
S. Capozziello  and  V. Faraoni, {\it Beyond Einstein Gravity}, Fundamental Theories of Physics Vol. 170, Springer Ed., Dordrecht  (2011).
 
 \bibitem{Chimento:2003iea}
  L.~P.~Chimento, A.~S.~Jakubi, D.~Pavon and W.~Zimdahl, 
  Phys.\ Rev.\ D {\bf 67} (2003) 083513 

\bibitem{Olive} 
  A. Coc, K.A. Olive, J.P. Uzan and E.Vangioni, [arXiv:0811.1845 [astro-ph]].

  
\bibitem{Gilles} 
  G. Esposito-Far\`ese, D. Polarski, {\it Phys. Rev.} {\bf D63}, 063504 (2001).

\bibitem{Clarkson-Ananda}
 C.~Clarkson, K.~Ananda and J.~Larena,  
  Phys.\ Rev.\ D {\bf 80} (2009) 083525; 
  T.~Buchert and S.~R�s�nen,   
  Ann.\ Rev.\ Nucl.\ Part.\ Sci.\  {\bf 62} (2012) 57. 


\bibitem{Marozzi_Luminosity_redshift_relation} 
  I.~Ben-Dayan, M.~Gasperini, G.~Marozzi, F.~Nugier and G.~Veneziano, 
  JCAP {\bf 1204} (2012) 036; 
  JCAP {\bf 1306} (2013) 002; 
  JCAP {\bf 1211} (2012) 045. 



\bibitem{Ellis_Stoeger}
G. F. R. Ellis and W. Stoeger, 
Class. Quant. Grav. {\bf 4} (1987) 1697.


\bibitem{Buchert:1995fz}   T.~Buchert and J.~Ehlers, 
  Astron.\ Astrophys.\  {\bf 320} (1997) 1. 

\bibitem{Buchert:1999er}
  T.~Buchert, 
  Gen.\ Rel.\ Grav.\  {\bf 32} (2000) 105. 
  
  
  \bibitem{Marozzi_gauge_invariant}
  M.~Gasperini, G.~Marozzi and G.~Veneziano,   
  JCAP {\bf 0903} (2009) 011; 
  JCAP {\bf 1002} (2010) 009; 
 %
 G.~Marozzi,   
 JCAP {\bf 1101} (2011) 012 
 
 
\bibitem{Buchert:2001sa}
  T.~Buchert,
  Gen.\ Rel.\ Grav.\  {\bf 33} (2001) 1381
  [gr-qc/0102049].
 
\bibitem{Reviews}
  G.~F.~R.~Ellis and T.~Buchert, 
  Phys.\ Lett.\ A {\bf 347} (2005) 38;  
%
  S.~Rasanen, 
  JCAP {\bf 0611} (2006) 003;  
   arXiv:1012.0784 [astro-ph.CO]; 
%
  T.~Buchert,   
  Gen.\ Rel.\ Grav.\  {\bf 40} (2008) 467;  
%
  T.~Clifton,   
  IJMPD {\bf 22}, 133004 (2013) 




\bibitem{Chris-George-Obinna} 
  C.~Clarkson, G.~Ellis, J.~Larena and O.~Umeh,
  Rept.\ Prog.\ Phys.\  {\bf 74} (2011) 112901. 

\bibitem{refs_Non_Newtonian}
  S.~Rasanen, 
  Phys.\ Rev.\ D {\bf 81} (2010) 103512; 
  E.~W.~Kolb, S.~Matarrese and A.~Riotto,   
  astro-ph/0511073;
  New J.\ Phys.\  {\bf 8} (2006) 322;  
  G.~F.~R.~Ellis and P.~K.~S.~Dunsby,   
  Astrophys.\ J.\  {\bf 479} (1997) 97; 
  J.~M.~M.~Senovilla, C.~F.~Sopuerta and P.~Szekeres,   
  Gen.\ Rel.\ Grav.\  {\bf 30} (1998) 389; 
  P.~Szekeres and T.~Rainsford,   
  Gen.\ Rel.\ Grav.\  {\bf 32} (2000) 479; 
  P.~Szekeres,   
  Gen.\ Rel.\ Grav.\  {\bf 32} (2000) 1025.

\bibitem{Rasanen_July2011}
  S.~Rasanen,   
  Phys.\ Rev.\ D {\bf 85} (2012) 083528. 
  



   \bibitem{Backreaction_simple_models}
  S.~Rasanen, 
  JCAP {\bf 0611} (2006) 003 
%
  C.~-H.~Chuang, J.~-A.~Gu and W-Y.~P.~Hwang, 
  Class.\ Quant.\ Grav.\  {\bf 25} (2008) 175001
%
  A.~Paranjape and T.~P.~Singh, 
  Class.\ Quant.\ Grav.\  {\bf 23} (2006) 6955 
%
  T.~Kai, H.~Kozaki, K.~-i.~nakao, Y.~Nambu and C.~-M.~Yoo,  
  Prog.\ Theor.\ Phys.\  {\bf 117} (2007) 229 
  %
  
  \bibitem{Rasanen_varia}
  C. Boehm and S.~Rasanen, 1305.7139v1; 
  S.~Rasanen,   
  JCAP {\bf 1003} (2010) 018; 
  JCAP {\bf 0902} (2009) 011. 

   
    
\bibitem{Rockhee_et_al}  
C.~Clarkson, T.~Clifton, A.~Coley and R.~Sung, 
  Phys.\ Rev.\ D {\bf 85}, 043506 (2012);  
  M.~Seikel and D.~J.~Schwarz,   
  arXiv:0912.2308 [astro-ph.CO];  
  P.~Fleury, H�l�n.~Dupuy and J.~-P.~Uzan,   
  Phys.\ Rev.\ D {\bf 87} (2013) 123526; 
  K.~Bolejko and P.~G.~Ferreira, 
  JCAP {\bf 1205} (2012) 003 


\bibitem{Paranjape:2009zu}
  A.~Paranjape,   
  arXiv:0906.3165 [astro-ph.CO].

\bibitem{Li-Thesis}     N.~Li, M.~Seikel and D.~J.~Schwarz,   
  Fortsch.\ Phys.\  {\bf 56} (2008) 465; 
  N.~Li and D.~J.~Schwarz,   
  Phys.\ Rev.\ D {\bf 76} (2007) 083011; 
  Phys.\ Rev.\ D {\bf 78} (2008) 083531.

  

\bibitem{Zalaletdinov1-2}
  R.~M.~Zalaletdinov,   
  Bull.\ Astron.\ Soc.\ India {\bf 25}, 401 (1997);  
  Int.\ J.\ Mod.\ Phys.\ A {\bf 23}, 1173 (2008). 

\bibitem{Wiltshire1-2}
  D.~L.~Wiltshire,   
  Phys.\ Rev.\ Lett.\  {\bf 99} (2007) 251101;  
  New J.\ Phys.\  {\bf 9} (2007) 377. 

\bibitem{Vitagliano:2009zy}
  V.~Vitagliano, S.~Liberati and V.~Faraoni,
  Class.\ Quant.\ Grav.\  {\bf 26} (2009) 215005 
  


\bibitem{vorticity}
  T.~H.~-C.~Lu, K.~Ananda, C.~Clarkson and R.~Maartens,
  JCAP {\bf 0902} (2009) 023
  [arXiv:0812.1349 [astro-ph]].
  A.~J.~Christopherson and K.~A.~Malik,
  Class.\ Quant.\ Grav.\  {\bf 28} (2011) 114004
  [arXiv:1010.4885 [gr-qc]].

\bibitem{Ellis-Cosmological Models}
P. K. S. Dunsby, M. Bruni M and G. F. R. Ellis  {\it Astrophys. J.}  {\bf 395} 54 (1992); 
  P.~K.~S.~Dunsby,
  Phys.\ Rev.\ D {\bf 48}, 3562 (1993);
  H.~van Elst and C.~Uggla,   
  Class.\ Quant.\ Grav.\  {\bf 14} (1997) 2673;   
  G.~F.~R.~Ellis and H.~van Elst,   
  NATO Adv.\ Study Inst.\ Ser.\ C.\ Math.\ Phys.\ Sci.\  {\bf 541} (1999) 1 
  



\bibitem{Gourgoulhon:2007ue}
  E.~Gourgoulhon,
  gr-qc/0703035 [GR-QC].
  


\bibitem{movingDE}
  A.~L.~Maroto,  
  JCAP {\bf 0605} (2006) 015;  
  Int.\ J.\ Mod.\ Phys.\ D {\bf 15} (2006) 2165; 
  AIP Conf.\ Proc.\  {\bf 878} (2006) 240; 
  J.~Beltran Jimenez and A.~L.~Maroto,
  Phys.\ Rev.\ D {\bf 76} (2007) 023003;  
  JCAP {\bf 0903} (2009) 015 
  T.~Harko and F.~S.~N.~Lobo,  
  JCAP {\bf 1307} (2013) 036 

  
\bibitem{Ballesteros:2010ks}
  G.~Ballesteros and J.~Lesgourgues,   
  JCAP {\bf 1010} (2010) 014.  


\bibitem{Sawicki:2012re}
  I.~Sawicki, I.~D.~Saltas, L.~Amendola and M.~Kunz,
  JCAP {\bf 1301} (2013) 004. 

  
\bibitem{Buchert:2002ij}
  T.~Buchert and M.~Carfora,   
  Phys.\ Rev.\ Lett.\  {\bf 90} (2003) 031101, 

  \bibitem{DE_Marozzi}
  I.~Ben-Dayan, M.~Gasperini, G.~Marozzi, F.~Nugier and G.~Veneziano,  
  Phys.\ Rev.\ Lett.\  {\bf 110} (2013) 021301 

\bibitem{Faraoni:2006fx} 
V.~Faraoni, E.~Gunzig and P.~Nardone,   
  Fund.\ Cosmic Phys.\  {\bf 20}, 121 (1999); 
  V.~Faraoni and S.~Nadeau,  
  Phys.\ Rev.\ D {\bf 75}, 023501 (2007); 
  D.~I.~Kaiser,   
  Phys.\ Rev.\ D {\bf 81}, 084044 (2010). 


\bibitem{Umeh:2010pr} 
  O.~Umeh, J.~Larena and C.~Clarkson,
  JCAP {\bf 1103}, 029 (2011)
  [arXiv:1011.3959 [astro-ph.CO]].


\bibitem{Sokolowski:1989ee} 
  L.~M.~Sokolowski,   
  Class.\ Quant.\ Grav.\  {\bf 6}, 2045 (1989).

\bibitem{Magnano:1993bd} 
  G.~Magnano and L.~M.~Sokolowski,  
  Phys.\ Rev.\ D {\bf 50}, 5039 (1994)
 


\bibitem{Ballesteros:2012kv}
  G.~Ballesteros and B.~Bellazzini,   
  JCAP {\bf 1304} (2013) 001  

  \bibitem{Ballesteros:2013nwa}
  G.~Ballesteros, B.~Bellazzini and L.~Mercolli,   
  arXiv:1312.2957 [hep-th].


  \bibitem{Faraoni_Review} 
  V. Faraoni, E. Gunzig and P. Nardone, Fund.Cosmic Phys. {\bf 20} (1999) 121.

  



\end{thebibliography}
\end{document}